\documentclass[11pt,a4paper]{article}
\usepackage{jheppub}
\usepackage[T1]{fontenc}
\usepackage{fix-cm}
\usepackage{lmodern}
\usepackage{amsmath,amssymb,mathtools,mathrsfs,empheq}
\usepackage{physics}
\usepackage{bbm}
\usepackage{bm}
\usepackage{hhline}
\usepackage{cancel}

\newcommand\nn{\nonumber}

\newcommand\mX{\mathcal{X}}

\newcommand\tsigma{\tilde{\sigma}}
\newcommand\mfa{\mathfrak{a}}
\newcommand\mfb{\mathfrak{b}}
\newcommand\mfc{\mathfrak{c}}

\newcommand\tnabla{\widetilde{\nabla}}

\newcommand\bg{\bar{g}}
\newcommand\bh{\bar{h}}
\newcommand\tir{\tilde{r}}

\newcommand\mfg{\mathfrak{g}}
\newcommand\bK{\overline{K}}
\newcommand\bmR{\overline{\mR}}
\newcommand\tmu{\tilde{\mu}}
\newcommand\bs{\bar{s}}
\newcommand\mI{\mathcal{I}}

\newcommand\bp{\bar{p}}
\newcommand\bq{\bar{q}}

\newcommand\Vol{\text{Vol}}

\newcommand\hA{\widehat{A}}

\newcommand\mta{\mathtt{a}}
\newcommand\mtb{\mathtt{b}}
\newcommand\mtc{\mathtt{c}}

\newcommand\onabla{\overline{\nabla}}

\makeatletter
\newcommand*{\rom}[1]{\expandafter\@slowromancap\romannumeral #1@}
\makeatother

\def\ri{{\rm i}}

\usepackage{tikz}

\usepackage{xcolor}

\newcommand\fft[2]{\frac{#1}{#2}}

\def\SO{{\rm SO}}

\def\SU{{\rm SU}}
\def\U{{\rm U}}

\usepackage{tensor}

\newcommand\mR{\mathcal{R}}

\newcommand\mN{\mathcal{N}}
\newcommand\mO{\mathcal{O}}
\newcommand\mA{\mathcal{A}}

\newcommand\mV{\mathcal{V}}

\newcommand\mY{\mathcal{Y}}
\newcommand\mS{\mathcal{S}}

\title{AdS$_7$ Black Holes and Holography}

\author[a,b]{Nikolay Bobev,}
\author[a,b]{Marina David,}
\author[c,d]{Vasil Dimitrov,}
\author[e]{and Junho Hong}

\affiliation[a]{Institute for Theoretical Physics, KU Leuven\,,\\ Celestijnenlaan 200D, B-3001 Leuven, Belgium}

\affiliation[b]{Leuven Gravity Institute, KU Leuven\,,\\ Celestijnenlaan 200D, B-3001 Leuven, Belgium}

\affiliation[c]{Dipartimento di Matematica, Universit\`a di Torino, Via Carlo Alberto 10, 10123 Torino, Italy}  
\affiliation[d]{INFN, Sezione di Torino, Via Pietro Giuria 1, 10125 Torino, Italy}

\affiliation[e]{Department of Physics \& Center for Quantum Spacetime, Sogang University\,,\\ 35 Baekbeom-ro, Mapo-gu, Seoul 04107, Republic of Korea}

\emailAdd{nikolay.bobev@kuleuven.be}
\emailAdd{marina.david@kuleuven.be}
\emailAdd{vasilradoslavov.dimitrov@unito.it}
\emailAdd{junhohong@sogang.ac.kr}

\abstract{We present a detailed study of the regularized on-shell action of the recently found asymptotically AdS$_7$ black hole supergravity solution with three angular momenta and two electric charges. We show that a particular choice of finite counterterms in the holographic renormalization procedure yields a result for the on-shell action of the supersymmetric limit of the black hole solution which is in perfect agreement with the large $N$ limit of the superconformal index of the dual 6d $\mathcal{N}=(2,0)$ $A_{N}$ SCFT on $S^1\times S^5$. We also discuss a generalization of this 7d supergravity background to a new general family of black holes with horizons given by the $L^{p,q,r}$ 5d Sasaki-Einstein manifolds. The regularized on-shell action of these new 7d black holes is also in agreement with supersymmetric localization results for the 6d $\mathcal{N}=(2,0)$ SCFT on $S^1\times L^{p,q,r}$. Finally, we briefly discuss a generalization of known AdS$_5$ gauged supergravity backgrounds to new 5d black holes with $L(p,q)$ lens space horizon topologies.}

\begin{document}
	
\maketitle 


\section{Introduction}\label{sec:intro}

Black hole solutions in AdS are interesting in their own right and are also of special importance for holography, since they provide insights into the thermal physics of the dual strongly interacting CFT. This is especially true for isolated CFTs without a weakly coupled description. A prime example in this class is the 6d $\mathcal{N}=(2,0)$ SCFT arising on the worldvolume of $N$ coincident M5-branes in M-theory. This holographic system strongly motivates the study of asymptotically AdS$_7$ black hole solutions in M-theory and is the subject of the current work.

The 6d conformal invariance and the $\SO(5)$ R-symmetry of the SCFT suggest that the general black hole solution in the dual supergravity description should be characterized by its mass, along with three independent angular momenta and two electric charges. Indeed, in our previous work~\cite{Bobev:2023bxl} we constructed such a solution of M-theory by exploiting the consistent truncation of 11d supergravity on $S^4$ to the maximal 7d gauged supergravity and a further consistent truncation to the $\U(1) \times \U(1)$ invariant sector of the 7d theory, which we refer to as the BDHM black hole. The supersymmetric limit of the solution in~\cite{Bobev:2023bxl} can be successfully compared to the large $N$ limit of the dual SCFT by showing that the supersymmetric partition function of the SCFT on $S^1\times S^5$, to leading order in the large $N$ limit, agrees with the on-shell action of the Euclidean supergravity background. This non-trivial precision test of the holographic duality was shown in~\cite{Bobev:2023bxl} by employing a sleight of hand. The supergravity on-shell action was evaluated by first computing the mass, entropy, angular momenta, and electric charges of the black hole solution and then relying on the integrated form of the first law of black hole thermodynamics known as the Quantum Statistical Relation (QSR). Our goal here is to derive this supergravity on-shell action by more direct methods and both demonstrate the validity of the first law of black hole thermodynamics and test the holographic correspondence.

To evaluate the on-shell action of interest we resort to the well-tested holographic renormalization method. This procedure provides the required covariant counterterms built out of the boundary fields of an asymptotically locally AdS$_7$ background which render any on-shell action of a given gravitational theory finite. Implementing this algorithmic procedure quickly leads to two challenges. The BDHM solution in~\cite{Bobev:2023bxl} has a fairly complicated form which results in unwieldy expressions when evaluating the bulk on-shell action and the necessary boundary counterterms. We tackle this problem with a judicious use of the equations of motion combined with symbolic calculations in \texttt{Mathematica}. On the more conceptual side we are faced with the problem of selecting a proper holographic renormalization scheme which is ambiguous due to the presence of covariant counterterms that can be built out of the boundary metric. We study these finite counterterms in some detail and show that there is a simple choice which yields a result for the on-shell action that is both compatible with the QSR and agrees with the supersymmetric partition function of the dual holographic SCFT as described in~\cite{Bobev:2023bxl}. We therefore propose this holographic renormalization scheme as the correct procedure for the evaluation of the on-shell action of asymptotically AdS$_7$ black holes in 7d supergravity.

The BDHM solution with three equal angular momenta can be presented in a suggestive form in terms of a squashed metric on $S^5$ written as a $\U(1)$ bundle over the K\"ahler manifold $\mathbb{CP}^2$. We use this to find a generalization of the BDHM background for which the horizon is given by the family of toric Sasaki-Einstein manifolds $L^{p,q,r}$ found in~\cite{Cvetic:2005ft,Cvetic:2005vk}. We calculate the mass, entropy, charges, and angular momentum of this new class of asymptotically AdS$_7$ black holes, as well as their on-shell actions computed via holographic renormalization. We verify the first law of black hole thermodynamics and discuss their holographic interpretation as a supergravity dual description of the 6d $\mathcal{N}=(2,0)$ SCFT placed on $S^1\times L^{p,q,r}$. In addition, we present a conjecture for the large $N$ limit of the supersymmetric partition function of the 6d SCFT on this manifold for general angular and electric fugacities and test our proposal in various limits by making contact with available results from supersymmetric localization.

Finally, inspired by the 7d black hole with Sasaki-Einstein horizons, we revisit the known 5d black hole solutions of the STU model in 5d gauged supergravity with $S^3$ horizons. We show that these backgrounds admit a simple generalization to asymptotically locally AdS$_5$ black holes with $L(p,q)$ lens space horizons and $\mathbb{R}\times L(p,q)$ conformal boundary. We also briefly discuss the holographic interpretation of these supergravity backgrounds.

We continue in the next section with a summary of the truncation of the 7d maximal gauged supergravity of interest and the BDHM solution. We then proceed to evaluate the on-shell action of this solution in Section~\ref{sec:OSA} and discuss in detail the procedure of holographic renormalization and choice of finite counterterms. In Section~\ref{sec:newBH} we generalize the BDHM solution with equal rotation parameters to a large family of 7d black holes with toric Sasaki-Einstein horizons given by the $L^{p,q,r}$ manifolds. In Section~\ref{sec:newBH-5} we shift gears to 5d and briefly describe a generalization of the known three-charge, two-angular momentum solution of the STU model in 5d gauged supergravity to a black hole with $L(p,q)$ lens space horizon, focusing on the equal-rotation limit. Section~\ref{sec:discussion} is devoted to a short conclusion and a discussion of some open questions. In the appendices we summarize our conventions and present some of the technical details pertaining to the evaluation of the on-shell action.
	
\section{7d gauged supergravity and the AdS$_7$ black hole}\label{sec:setup}
In this section, we provide a concise overview of the 7d gauged supergravity theory relevant to our work -- specifically, the $\U(1)\times \U(1)$ truncation of maximal $\SO(5)$ gauged supergravity -- and the AdS$_7$ black hole solution found in~\cite{Bobev:2023bxl}.

\subsection{$\U(1)\times \U(1)$ truncation of 7d gauged supergravity}\label{sec:setup:sugra}

The BDHM black hole solution resides within a $\U(1)\times \U(1)$ invariant sector~\cite{Liu:1999ai} of 7d maximal $\SO(5)$ gauged supergravity, which in turn arises as a consistent truncation of 11d supergravity on $S^4$~\cite{Nastase:1999cb,Nastase:1999kf}. The bosonic sector of the truncated theory comprises the 7d metric, two Abelian gauge fields, two real scalar fields, and a single 3-form potential with a ``self-dual'' 4-form field strength. A more detailed presentation of the full 7d maximal gauged supergravity can be found in~\cite{Pernici:1984xx,Liu:1999ai,Gauntlett:2000ng}.

\medskip

For completeness, we present the bosonic action for the $\U(1)\times \U(1)$ truncated 7d theory~\cite{Chow:2007ts,Bobev:2023bxl},
\begin{align}
	S_\text{bulk}^L&=\fft{1}{16\pi G_N}\int\bigg[\star(R +2\mfg^2\mV) -\fft12\sum_{I=1}^2d\varphi_I\wedge \star d\varphi_I-\fft12\sum_{I=1}^2\fft{1}{X_I^2}F^I_{(2)}\wedge \star F^{I}_{(2)}\nn\\
	&\kern6em~-\fft{1}{2}X_1^2X_2^2F_{(4)}\wedge \star F_{(4)}+\mfg F_{(4)}\wedge A_{(3)}+F^1_{(2)}\wedge F^2_{(2)}\wedge A_{(3)}\bigg]\,,\label{S:L}
\end{align}
where the superscript ``$L$'' stands for (mostly plus) Lorentzian signature. In the expression above we have used the following scalar parametrization and the field strengths 
\begin{equation}
	X_1 = e^{-\frac{1}{\sqrt{10}}\varphi_1-\frac{1}{\sqrt{2}}\varphi_2}\,, \quad X_2 = e^{-\frac{1}{\sqrt{10}}\varphi_1+\frac{1}{\sqrt{2}}\varphi_2}\,, \quad F_{(2)}^{I} = dA_{(1)}^{I}\,, \quad F_{(4)} = dA_{(3)}\,.
\end{equation}
The real parameter $\mfg$ is the gauge coupling in the supergravity theory and the scalar potential takes the form
\begin{align}
	\mV&=8X_1X_2+\fft{4(X_1+X_2)}{X_1^2X_2^2}-\fft{1}{X_1^4X_2^4}\,.
\end{align}
Our conventions for the Hodge star operator are specified in Appendix \ref{app:conventions}.

\medskip

From this action, one can derive the following equations of motion. The Einstein equations take the form  
\begin{subequations}\label{eom}
\begin{equation}
	\begin{split}
		0&=R_{\mu\nu}-\fft12g_{\mu\nu}\big[R+2\mfg^2\mV\big]\\
		&\quad-\sum_{I=1}^2\bigg[\fft12\partial_\mu\varphi_I\partial_\nu\varphi_I-\fft14g_{\mu\nu}\partial^\rho\varphi_I\partial_\rho\varphi_I\bigg]-\sum_{I=1}^2\fft{1}{X_I^2}\bigg[\fft12F^I_{\mu\rho}F^I_{\nu}{}^\rho-\fft18g_{\mu\nu}F^I_{\rho\sigma}F^{I\,\rho\sigma}\bigg]\\
		&\quad-X_1^2X_2^2\bigg[\fft{1}{12}F_{\mu\rho\sigma\lambda}F_{\nu}{}^{\rho\sigma\lambda}-\fft{1}{96}g_{\mu\nu}F_{\rho\sigma\lambda\delta}F^{\rho\sigma\lambda\delta}\bigg]\,.
	\end{split}\label{eom:Einstein}
\end{equation}
The scalar field equations are  
\begin{equation}
	\begin{split}
		\nabla^\mu\nabla_\mu\varphi_1&=\fft{1}{2\sqrt{10}}\sum_{I=1}^2\fft{1}{X_I^2}F^I_{\mu\nu}F^{I\,\mu\nu}-\fft{1}{12\sqrt{10}}X_1^2X_2^2F_{\mu\nu\rho\sigma}F^{\mu\nu\rho\sigma}\\
		&\quad+\fft{8\mfg^2}{\sqrt{10}}\bigg(4X_1X_2-\fft{3(X_1+X_2)}{X_1^2X_2^2}+\fft{2}{X_1^4X_2^4}\bigg)\,,\\
		\nabla^\mu\nabla_\mu\varphi_2&=\fft{1}{2\sqrt{2}}\bigg(\fft{1}{X_1^2}F^1_{\mu\nu}F^{1\,\mu\nu}-\fft{1}{X_2^2}F^2_{\mu\nu}F^{2\,\mu\nu}\bigg)+4\sqrt{2}\mfg^2\fft{X_1-X_2}{X_1^2X_2^2}\,,
	\end{split}\label{eom:scalar}
\end{equation}
while the equations of motion for the gauge fields read 
\begin{equation}
\begin{split}
	d(X_1^{-2} \star F^1_{(2)})&=F^2_{(2)}\wedge F_{(4)}\,,\\
	d(X_2^{-2} \star F^2_{(2)})&=F^1_{(2)}\wedge F_{(4)}\,.
\end{split}\label{eom:vector}
\end{equation}
The equation of motion for the 3-form potential is 
\begin{align}
	d(X_1^2X_2^2 \star F_{(4)})&=2\mfg F_{(4)}+F^1_{(2)}\wedge F^2_{(2)}\,.\label{eom:3form}
\end{align}
\end{subequations}
In addition, the 3-form potential $A_{(3)}$ is subject to a ``self-duality'' condition~\cite{Pilch:1984xy} 
\begin{equation}
	X_1^2X_2^2 \star F_{(4)}=2\mfg A_{(3)}+\fft12\left(A^1_{(1)}\wedge F^2_{(2)}+A^2_{(1)}\wedge F^1_{(2)}\right)-dA_{(2)}\,,\label{self-dual}
\end{equation}
where $A_{(2)}$ is an auxiliary 2-form potential introduced to impose the self-duality. The existence of such an $A_{(2)}$ can be demonstrated locally by applying the Poincar\'e lemma to the 3-form field equation \eqref{eom:3form}. Consequently, the self-duality condition \eqref{self-dual} does not impose any additional independent constraints on local solutions of the theory.

\subsection{BDHM AdS$_7$ black hole}\label{sec:setup:bdhm}
We now continue with the presentation of the BDHM solution which is a non-extremal non-supersymmetric asymptotically AdS$_7$ black hole, characterized by three independent angular momenta and two electric charges~\cite{Bobev:2023bxl}. 

\medskip
 
The metric for this solution in the Boyer-Lindquist type coordinate system reads
\begin{align}
	ds^2&= (H_1H_2)^{\fft15}\Bigg[-(1+\mfg^2r^2)\sum_{i=1}^3\fft{\mu_i^2}{\Xi_i}dt^2+\fft{\sum_{i=1}^3\fft{\mu_i^2}{r^2+a_i^2}\prod_{j=1}^3(r^2+a_j^2)}{V(r)-2mW(r)}dr^2 \nn \\
	&\kern5em~+\sum_{i=1}^3\fft{r^2+a_i^2}{\Xi_i}d\mu_i^2-\fft{\mfg^2}{(1+\mfg^2r^2)\sum_{i=1}^3\fft{\mu_i^2}{\Xi_i}}\bigg(\sum_{j=1}^3\fft{r^2+a_j^2}{\Xi_j}\mu_j d\mu_j\bigg)^2  \nn \\
	&\kern5em~+\sum_{i=1}^3\fft{r^2+a_i^2}{\Xi_i}\mu_i^2d\phi_i^2+\fft{1-\fft{1}{H_1}}{1-(s_2/s_1)^2}K_1^2+\fft{1-\fft{1}{H_2}}{1-(s_1/s_2)^2}K_2^2 \Bigg]\,.\label{BDHM}
\end{align}
The various parameters and functions used above are defined as 
\begin{subequations}
	\begin{align}
		\mu_i&=\sqrt{\fft{(a_i^2-y^2)(a_i^2-z^2)}{\prod_{j=1\,(\neq i)}^3(a_i^2-a_j^2)}}\qquad\Big(\sum_{i=1}^3\mu_i^2=1\Big)\,,\\
		s_I&=\sinh\delta_I\,,\qquad c_I=\cosh\delta_I\,,\qquad \Xi_i = 1-a_i^2\mfg^2\,, \\
		H_I&=1+\fft{2ms_I^2}{\sum_{i=1}^3\fft{\mu_i^2}{r^2+a_i^2}\prod_{j=1}^3(r^2+a_j^2)}=1+\fft{2ms_I^2}{(r^2+y^2)(r^2+z^2)}\,,\\
		V(r)&=\fft{(1+\mfg^2r^2)\prod_{i=1}^{3}(r^2+a_i^2)}{r^2}\,,\\
		W(r)&=1-\fft12(s_1^2+s_2^2)(2\mfg^2r^2 + \Sigma_2) -\fft{2m\mfg^2s_1^2s_2^2}{r^2}+\fft{(s_1^2+s_2^2)\Pi_1}{\mfg^2r^2}  \\
		&\quad-
		\fft{(c_1-c_2)^2}{\mfg^2r^2}\prod_{i=1}^3\bigg(a_i\mfg+\fft{\Pi_1}{a_i\mfg}\bigg)-\fft{(c_1-c_2)^2}{4}\bigg(2\Sigma_2+8\Pi_1+\Sigma_1\prod_{i=1}^3(\Sigma_1-2a_i\mfg)\bigg)\,,\nn\\
		K_1&=\fft{c_1+c_2}{2s_1}\mX+\fft{c_1-c_2}{2s_1}\mY\,,\qquad K_2=\fft{c_1+c_2}{2s_2}\mX-\fft{c_1-c_2}{2s_2}\mY\,,\label{1-forms}\\
		\mX&=\sum_{i=1}^3\fft{\mu_i^2}{\Xi_i}(dt-a_id\phi_i)\,,\\
		\mY&=\sum_{i=1}^3\fft{\mu_i^2}{\Xi_i}\bigg[(1-\Sigma_2-2\Pi_1)dt+a_i\bigg(1+\Sigma_2-2a_i^2\mfg^2+\fft{2\Pi_1}{a_i^2\mfg^2}\bigg)d\phi_i\bigg]\,.
	\end{align}
\end{subequations}
We also define the following quantities, which are used throughout the manuscript, for notational convenience:
\begin{align}
	\Sigma_n&\equiv(a_1\mfg)^n+(a_2\mfg)^n+(a_3\mfg)^n\,, \nn\\
	\Pi_n&\equiv(a_1\mfg)^n(a_2\mfg)^n(a_3\mfg)^n\,,\label{SigmaPi}\\
	\Pi_{nm}&\equiv\sum_{i=1}^3(a_i\mfg)^n\big(\Sigma_m-(a_i\mfg)^m\big)\,.\nn
\end{align}
The coordinate ranges are given by 
\begin{equation}
	r_+\leq r\,,\qquad 0\leq a_1\leq z\leq a_2\leq y\leq a_3~~(0\leq\mu_i\leq 1)\,,\qquad 0\leq\phi_{1,2,3}<2\pi\,,\label{BH:range}
\end{equation}
where $r_+$ is the largest positive real root of the equation $V(r)-2mW(r)=0$ and denotes the outer horizon radius of the black hole.

The two scalar fields are expressed as
\begin{equation}
	X_I=\fft{(H_1H_2)^\fft25}{H_I}\,.\label{ansatz:scalar}
\end{equation}

The two Abelian gauge fields can be written as the following 1-forms
\begin{equation}\label{eq:A1Idef}
	A^I_{(1)}=\bigg(1-\fft{1}{H_I}\bigg)K_I \,,
\end{equation}
where we focus on a local solution and omit the flat connection required for regular behavior near the horizon discussed in~\cite{Bobev:2023bxl}. This aspect will be addressed in the next subsection.

The 3-form reads
\begin{equation}
	\begin{split}
		A_{(3)}&=2ms_1s_2\mfg^4a_1a_2a_3\bigg[\mA[y^2,z^2,0]-\mA[y^2,z^2,\mfg^{-2}]\bigg]\\
		&\hspace{3mm}\wedge\bigg[\fft{dz\wedge\left(\mA[y^2,0,0]-\mA[y^2,0,\mfg^{-2}]\right)}{(r^2+y^2)z}+\fft{dy\wedge\left(\mA[z^2,0,0]-\mA[z^2,0,\mfg^{-2}]\right)}{(r^2+z^2)y}\bigg]\\
		&\hspace{3mm}+2ms_1s_2\mfg^3\mA[y^2,z^2,0]\\
		&\hspace{3mm}\wedge\left[\fft{z \, dz\wedge\left(\mA[y^2,0,0]-\mA[y^2,0,\mfg^{-2}]\right)}{(r^2+y^2)}+\fft{y\, dy\wedge\left(\mA[z^2,0,0]-\mA[z^2,0,\mfg^{-2}]\right)}{(r^2+z^2)}\right]\,,\label{ansatz:3form}
	\end{split}
\end{equation}
where we have defined
\begin{equation}
	\mA[v_1,v_2,v_3]=\Bigg[\prod_{i=1}^3\fft{1-\mfg^2v_i}{\Xi_i}\Bigg]dt-\sum_{i=1}^3\Bigg[\fft{(a_i^2-v_1)(a_i^2-v_2)(a_i^2-v_3)}{a_i\Xi_i\prod_{j=1\,(\neq i)}^3(a_i^2-a_j^2)}\Bigg]d\phi_i\,.
\end{equation}

The solution described above is specified by six parameters $(m,a_1,a_2,a_3, \delta_1,\delta_2)$, and satisfies the bosonic equations of motion (\ref{eom}) for the $\U(1)\times\U(1)$ truncation of 7d maximal gauged supergravity. The self-duality constraint \eqref{self-dual} can also be verified explicitly by employing the following 2-form potential
\begin{equation}
	\begin{split}
		A_{(2)}&=\bigg(\fft{1}{H_1}+\fft{1}{H_2}\bigg)\fft{ms_1s_2(a_1+a_2a_3\mfg)}{(r^2+y^2)(r^2+z^2)}\bigg(\fft{(1-\mfg^2y^2)(1-\mfg^2z^2)\mu_1^2}{\Xi_1(1-a_2^2\mfg^2)(1-a_3^2\mfg^2)}dt\wedge d\phi_1\\
		&\kern16em+\fft{\mfg(a_3^2-a_2^2)\mu_2^2\mu_3^2}{\Xi_2\Xi_3}d\phi_2\wedge d\phi_3\bigg)\\
		&\quad+\big(\text{cyclic-permutations}\big)\,,\label{ansatz:2form}
	\end{split}
\end{equation}
where the cyclic permutations are taken over the rotation parameters $a_{i}$ and the corresponding coordinates $\mu_i$ and $\phi_{i}$ for $i=1,2,3$. 

\subsection{Gauge shift}\label{sec:setup:gg}
To ensure that the BDHM solution described above represents a globally well-defined black hole background, one must impose appropriate regularity conditions. In particular, the 1-form and 3-form gauge potentials should be smooth near the horizon. This can be achieved by choosing a particular gauge in which the following components of these potentials vanish at the horizon as~\cite{Cabo-Bizet:2018ehj,Cassani:2019mms,Cassani:2022lrk,BenettiGenolini:2023rkq,BenettiGenolini:2023ucp}\footnote{In some cases, including the present one, regularity can alternatively be ensured by requiring that the norm of the gauge potential remains bounded at the horizon \cite{Bobev:2023bxl,Hong:2024uns}.} 
\begin{align}
	\ri_v \widehat{A}_{(1)}^I|_{r=r_+}=0\qquad\text{and}\qquad \ri_v \widehat{A}_{(3)}|_{r=r_+}=0\,,
\end{align}
where the hatted forms denote gauge-transformed potentials and $v$ is the Killing vector field corresponding to the null generator of the horizon. Below, we present the shifted gauge potentials consistent with this regularity condition.  

\medskip

The regularity of the gauge fields can be achieved by shifting the corresponding 1-forms by flat connections as
\begin{align}
	A_{(1)}^I\quad\to\quad \hA_{(1)}^I=A_{(1)}^I-\Phi_I dt\,,
\end{align}
where $A_{(1)}^I$ is defined in (\ref{eq:A1Idef}) and $\Phi^I$ denotes the electrostatic potentials given by~\cite{Bobev:2023bxl}
\begin{align}
	\Phi_1&=\fft{2mr_+^2s_1c_1\big(\prod_i(r_+^2+a_i^2)+2ms_2^2(r_+^2-a_1a_2a_3\mfg)\big)}{\mS(r_+)}\\
	&\quad-\fft{mr_+^2s_1(c_1-c_2)}{\mS(r_+)}\bigg\{(\Sigma_2+2\Pi_1)\big({\textstyle\prod_{i=1}^3}(r_+^2+a_i^2)+2ms_2^2(r_+^2-a_1a_2a_3\mfg)\big)\notag\\
	&\kern10em+4m\mfg s_2^2\prod_{i=1}^3\bigg(a_i+\fft{a_1a_2a_3\mfg}{a_i}\bigg)\bigg\}\,,\nn\\
	\mS(r)&\equiv \prod_{I=1}^2\Big(\textstyle{\prod_{i=1}^3}(r^2+a_i^2)+2ms_I^2(r^2-a_1a_2a_3\mfg)\Big)\\
	&\quad+2m\mfg(c_1-c_2)^2\prod_{i=1}^3(r^2+a_i^2)\bigg(a_i+\fft{a_1a_2a_3\mfg}{a_i}\bigg)\,.\nn
\end{align}
The electrostatic potential $\Phi_2$ associated with the second U(1) symmetry can be obtained by interchanging the charge parameters $\delta_1\leftrightarrow\delta_2$ in the expression above. 

\medskip

The 3-form potential can also be made regular near the horizon by implementing the gauge transformation 
\begin{align}
	A_{(3)}\quad\to\quad \hA_{(3)} = A_{(3)} + \fft{1}{4\mfg} d{t} \wedge \big(\Phi_1 F_{(2)}^2 + \Phi_2 F_{(2)}^1\big) - \fft{1}{2\mfg} d {A_{(2)}}\,,
\end{align}
where $A_{(3)}$ and $A_{(2)}$ stand for the expressions (\ref{ansatz:3form}) and (\ref{ansatz:2form}) respectively. 

\medskip

The self-duality condition (\ref{self-dual}) can be expressed in terms of the above gauge-shifted potentials as
\begin{equation}
	X_1^2X_2^2 \star F_{(4)}=2g\hA_{(3)}+\fft12\left(\hA^1_{(1)}\wedge F^2_{(2)}+\hA^2_{(1)}\wedge F^1_{(2)}\right)-d\hA_{(2)}\,,\label{self-dual:gg}
\end{equation}
where the auxiliary 2-form potential becomes pure gauge since $d\hA_{(2)}=0$. 

\medskip

After applying the above gauge transformations, the bosonic action of the $\U(1)\times \U(1)$ truncated 7d maximal gauged supergravity (\ref{S:L}) can be rewritten as
\begin{align}
	S_\text{bulk}^L&=\fft{1}{16\pi G_N}\int\bigg[\star(R +2\mfg^2\mV) -\fft12\sum_{I=1}^2d\varphi_I\wedge \star d\varphi_I-\fft12\sum_{I=1}^2\fft{1}{X_I^2}F^I_{(2)}\wedge \star F^{I}_{(2)}\nn\\
	&\kern6em~-\fft{1}{2}X_1^2X_2^2F_{(4)}\wedge \star F_{(4)}+\mfg F_{(4)}\wedge \hA_{(3)}+F^1_{(2)}\wedge F^2_{(2)}\wedge \hA_{(3)}\bigg]\,.\label{S:L:gg}
\end{align}
Although the final two Chern–Simons terms differ from their original expressions, the difference contributes only boundary terms and does not affect the bulk dynamics. Nevertheless, the form in (\ref{S:L:gg}) is particularly convenient for evaluating the Euclidean on-shell action via holographic renormalization, as we discuss in the following section.

\section{Euclidean on-shell action of the AdS$_7$ black hole}\label{sec:OSA}
In this section, we compute the regularized Euclidean on-shell action of the BDHM black hole reviewed in the previous section, following the well-established procedure of holographic renormalization~\cite{Bianchi:2001kw,Skenderis:2002wp}.

\subsection{Euclidean actions}\label{sec:OSA:action}
As a first step in evaluating the regularized Euclidean on-shell action of the BDHM background, we present the relevant bulk and boundary Euclidean actions that need to be computed on-shell.

\bigskip
\noindent\textbf{Bulk action}
\medskip

\noindent To obtain the 7d bulk action in Euclidean signature from its Lorentzian counterpart (\ref{S:L:gg}), we perform a Wick rotation 
\begin{align}
	t\quad\to\quad -\ri\tau\,.\label{Wick}
\end{align}
Adopting the Wick rotation convention detailed in Appendix~\ref{app:conventions}, along with the standard identification $S^E_\text{bulk}=-\ri S^L_\text{bulk}$, we obtain the Euclidean version of the bosonic action from the gauge-shifted Lorentzian form (\ref{S:L:gg})
\begin{align}
	S^E_\text{bulk}&=-\fft{1}{16\pi G_N}\int\bigg[\star(R +2\mfg^2\mV) -\fft12\sum_{I=1}^2d\varphi_I\wedge \star d\varphi_I-\fft12\sum_{I=1}^2\fft{1}{X_I^2}F^I_{(2)}\wedge \star F^{I}_{(2)}\nn\\
	&\kern7em-\fft{1}{2}X_1^2X_2^2F_{(4)}\wedge \star F_{(4)}+\ri \mfg F_{(4)}\wedge \widehat{A}_{(3)}+\ri F^1_{(2)}\wedge F^2_{(2)}\wedge \widehat{A}_{(3)}\bigg]\,.\label{S:E}
\end{align}
Note that the imaginary coefficients in the Chern–Simons terms arise from the Wick rotation. 

We now use the equations of motion reviewed in Section~\ref{sec:setup:sugra} to recast the Euclidean action~(\ref{S:E}) into a more convenient on-shell form. First, we eliminate the Einstein-Hilbert and scalar kinetic terms using the Einstein equations~(\ref{eom:Einstein}). Next, we apply the gauge field equations of motion~(\ref{eom:vector}) and the self-duality condition~(\ref{self-dual:gg}) to combine most of the terms into total derivatives, leading to the following expression
\begin{align}
	S_\text{bulk}^E\Big|_\text{(\ref{eom})} &= \fft{1}{16\pi G_N} \int \Bigg[ \fft{4\mfg^2}{5} \star \mV + \fft{\ri \mfg}{5} F_{(4)} \wedge \widehat{A}_{(3)} \notag \\
	& \quad+ d\,\bigg\{ \fft{1}{5} \sum_{I = 1}^{2}{} \hA_{(1)}^I \wedge X_I^{-2} \star F_{(2)}^I - \fft{\ri}{2} \big(\hA_{(1)}^1 \wedge F_{(2)}^2 + \hA_{(1)}^2 \wedge F_{(2)}^1\big) \wedge \hA_{(3)}\bigg\}\Bigg] \,, \label{S:E:onshell}
\end{align}
where the subscript ``(\ref{eom})'' indicates that the expression holds on-shell.

\bigskip
\noindent\textbf{Gibbons-Hawking-York term}
\medskip

\noindent To ensure a well-posed variational problem of the Einstein equations, the bulk action~(\ref{S:E}) must be supplemented by the Gibbons–Hawking–York (GHY) boundary term. In Euclidean signature, it takes the form
\begin{equation}
	S_\text{GHY}^E=-\fft{1}{8\pi G_N}\int d^6y\,\sqrt{h}\,K\,,\label{GHY}
\end{equation}
where the extrinsic curvature is defined as
\begin{equation}
	K_{\mu\nu}=\nabla_\mu n_\nu-n_\mu n^\rho\nabla_\rho n_\nu\,,
\end{equation}
in terms of an outward unit normal vector $n^\mu$ to the boundary. Here, $y$ collectively denotes the 6d coordinates of the boundary with the corresponding coordinate indices $\{i,j,k,l\}$ and $h_{ij}$ is the induced 6d metric.

\bigskip
\noindent\textbf{Counterterms}
\medskip

\noindent The on-shell values of the bulk action~\eqref{S:E:onshell} and the GHY term (\ref{GHY}) exhibit divergences for asymptotically locally EAdS$_7$ backgrounds. These divergences can be regularized via holographic renormalization~\cite{Bianchi:2001kw,Skenderis:2002wp}, by introducing a radial cutoff and adding appropriate local counterterms. For asymptotically locally EAdS$_7$ geometries, the required counterterms are given by, see for example~\cite{Bueno:2022log}, 
\begin{subequations}
	\begin{align}
		S^E_\text{ct}&=\sum_{\alpha=0}^3S^E_\text{ct-$\alpha$}\,,\\
		S^E_\text{ct-0}&=\fft{1}{8\pi G_N}\int d^6y\,\sqrt{h}\,5\mfg\,,\label{ct-inf:0der}\\
		S^E_\text{ct-1}&=\fft{1}{8\pi G_N}\int d^6y\,\sqrt{h}\,\fft{1}{8\mfg}\mR\,,\label{ct-inf:2der}\\
		S^E_\text{ct-2}&=\fft{1}{8\pi G_N}\int d^6y\,\sqrt{h}\,\fft{1}{64\mfg^3}\bigg[\mR_{ij}\mR^{ij}-\fft{3}{10}\mR^2\bigg]\,,\label{ct-inf:4der}\\
		S^E_\text{ct-3}&=\fft{\log\epsilon}{8\pi G_N}\int d^6y\,\sqrt{h}\,\fft{1}{128\mfg^5}\bigg[\mR\mR_{ij}\mR^{ij}-\fft{3}{25}\mR^3-2\mR_{ijkl}\mR^{ik}\mR^{jl}\nn\\
		&\kern12em+\fft25\mR^{ij}\tnabla_i\tnabla_j\mR-\mR^{ij}\tnabla^2\mR_{ij}+\fft{1}{10}\mR\tnabla^2\mR\bigg]\,,\label{ct-inf:log}
	\end{align}\label{ct-inf}%
\end{subequations}
where  $\mR_{ijkl}$, $\mR_{ij}$, and $\mR$ denote the Riemann tensor, Ricci tensor, and Ricci scalar on the boundary, respectively. The logarithmic term in $S^E_\text{ct-3}$ is associated with the holographic Weyl anomaly~\cite{Henningson:1998gx} and appears only in odd-dimensional gravitational backgrounds. We will elaborate on the radial cutoff parameter $\epsilon$ in Section~\ref{sec:OSA:frame}. 

While these counterterms cancel the divergences from the bulk and GHY actions, finite ambiguities remain. These are governed by the choice of additional finite counterterms, such as
\begin{subequations}
	\begin{align}
		S^E_\text{ct-fin}&=\sum_{x}\mtc_x S^E_\text{ct-fin-$x$}\,,\\
		S^E_\text{ct-fin-1}&=\fft{1}{8\pi \mfg^5G_N}\int d^6y\,\sqrt{h}\,\mR^3\,,\label{ct-fin:1}\\
		S^E_\text{ct-fin-2}&=\fft{1}{8\pi \mfg^5G_N}\int d^6y\,\sqrt{h}\,\mR\mR_{ij}\mR^{ij}\,,\\
		S^E_\text{ct-fin-3}&=\fft{1}{8\pi \mfg^5G_N}\int d^6y\,\sqrt{h}\,\mR_{ijkl}\mR^{ik}\mR^{jl}\,,\\
		S^E_\text{ct-fin-4}&=\fft{1}{8\pi \mfg^5G_N}\int d^6y\,\sqrt{h}\,\mR^{ij}\tnabla_i\tnabla_j\mR\,,\\
		S^E_\text{ct-fin-5}&=\fft{1}{8\pi \mfg^5G_N}\int d^6y\,\sqrt{h}\,\mR^{ij}\tnabla^2\mR_{ij}\,,\\
		S^E_\text{ct-fin-6}&=\fft{1}{8\pi \mfg^5G_N}\int d^6y\,\sqrt{h}\,\mR\tnabla^2\mR\,,
	\end{align}\label{ct-fin}%
\end{subequations}
with $\mtc_x$ arbitrary real constants. The list above is not exhaustive, as additional six-derivative combinations can be constructed from curvature tensors and covariant derivatives. A suitable prescription for choosing the coefficients of these finite counterterms will be discussed in Section~\ref{sec:OSA:holo}.

\bigskip
\noindent\textbf{Regularized Euclidean on-shell action}
\medskip

\noindent Combining the bulk action, the GHY boundary term, and both the infinite and finite counterterms presented above, the regularized Euclidean on-shell action for a given asymptotically locally EAdS$_7$ background $X$ reads
\begin{align}
	I_\text{reg}\Big|_X= \Big[S_\text{bulk}+S_\text{GHY}+S_\text{ct}+S_\text{ct-fin}\Big]_{X}\,,\label{I:reg}
\end{align}
where the subscript ``$X$'' indicates that the respective actions are evaluated on-shell for the background $X$. For notational convenience, we omit the superscript ``$E$'' indicating Euclidean signature from this point forward.

\subsection{Asymptotically non-rotating frame}\label{sec:OSA:frame}
Before evaluating the regularized on-shell action (\ref{I:reg}) for the BDHM black hole solution of interest, it is necessary to express the background in a coordinate system that admits a Fefferman–Graham (FG) expansion near the AdS boundary
\begin{align}
	ds^2=\fft{d\rho^2}{4\mfg^2\rho^2}+\fft{1}{\mfg^2\rho}g_{ij}(x,\rho)dx^idx^j\,,\label{FG}
\end{align}
which is essential for a proper application of holographic renormalization \cite{Bianchi:2001kw,Skenderis:2002wp}. Rather than working directly in FG coordinates, however, we adopt an alternative coordinate system following \cite{Gibbons:2004ai,Gibbons:2005vp,Gibbons:2005jd}, in which the BDHM solution asymptotically approaches the canonical global AdS$_7$ metric at large radius. Both coordinate systems describe geometries that are non-rotating in the asymptotic regime, ensuring that holographic renormalization can be consistently applied. 

\medskip

To guarantee that the BDHM background exhibits the desired asymptotic behavior described above, we perform a coordinate transformation from $(r,\mu_i)$ to $(\tir,\tmu_i)$ where the new coordinates are defined through \cite{Gibbons:2005vp,Gibbons:2005jd,Bobev:2023bxl}
\begin{align}
	\tmu_i^2=\fft{r^2+a_i^2}{\tir^2\Xi_i}\mu_i^2\qquad\text{s.t.}\qquad\sum_{i=1}^3\tmu_i^2=1\,,\label{coord:change}
\end{align}%
which implies the relation
\begin{align}
	\tir^2=\sum_{i=1}^3\fft{r^2+a_i^2}{\Xi_i}\mu_i^2 = \fft{(1+\mfg^2r^2)(1-\mfg^2y^2)(1-\mfg^2z^2)}{\mfg^2\Xi_1\Xi_2\Xi_3}-\fft{1}{\mfg^2}\,.\label{tir:ryz}
\end{align}
In terms of the new coordinates, and after performing the Wick rotation \eqref{Wick}, the BDHM metric (\ref{BDHM}) becomes
\begin{align}
	ds^2&=(H_1H_2)^{\fft15}\Bigg[(1+\mfg^2\tir^2)d\tau^2+\fft{d\tir^2}{1+\mfg^2\tir^2}+\tir^2\sum_{i=1}^3\Big(d\tmu_i^2+\tmu_i^2d\phi_i^2\Big)\nn\\
	&\quad+\fft{2m W(r)}{\tir^2\sum_{i=1}^3\fft{\Xi_i\tmu_i^2}{(r^2+a_i^2)^2}}\fft{1}{(1+\mfg^2r^2)(V(r)-2mW(r))}\bigg(\fft{d\tir}{\tir}+\tir^2\sum_{i=1}^3\fft{\Xi_i\tmu_i d\tmu_i}{r^2+a_i^2}\bigg)^2\nn\\
	&\quad+\fft{1-\fft{1}{H_1}}{1-(s_2/s_1)^2}K_1^2+\fft{1-\fft{1}{H_2}}{1-(s_1/s_2)^2}K_2^2\Bigg]\,,\label{BDHM:tilde}
\end{align}
where we retain the original radial coordinate $r$ in several places to avoid cumbersome expressions. This form makes it manifest that the Euclidean BDHM metric (\ref{BDHM:tilde}) approaches the global EAdS$_7$ metric at large radius
\begin{align}
	d\bar{s}^2&=(1+\mfg^2\tir^2)d\tau^2+\fft{d\tir^2}{1+\mfg^2\tir^2}+\tir^2ds_{S^5}^2\,, \quad\quad  ds_{S^5}^2\equiv\sum_{i=1}^3\Big(d\tmu_i^2+\tmu_i^2d\phi_i^2\Big)\,,\label{EAdS7}
\end{align}
where $ds_{S^5}^2$ denotes the unit 5-sphere metric and we have implicitly used the large $\tir$ expansions of the original coordinates
\begin{subequations}
	\begin{align}
		r^2&=(\sum_{i=1}^3\Xi_i\tmu_i^2)\tir^2-\fft{\sum_{j=1}^3a_j^2\Xi_j\tmu_j^2}{\sum_{i=1}^3\Xi_i\tmu_i^2}+\mO(\tir^{-2})\,,\\
		\mu_i^2&=\fft{\Xi_i\tmu_i^2}{\sum_{j=1}^3\Xi_j\tmu_j^2}\left[1+\fft{1}{\tir^2}\bigg(-\fft{a_i^2}{\sum_{j=1}^3\Xi_j\tmu_j^2}+\fft{\sum_{k=1}^3a_k^2\Xi_k\tmu_k^2}{(\sum_{j=1}^3\Xi_j\tmu_j^2)^2}\bigg)+\mO(\tir^{-4})\right]\,,
	\end{align}\label{tilde:expansion}%
\end{subequations}
determined from the coordinate transformation (\ref{coord:change}).

\medskip

One can also verify that the radial coordinate transformation $\tir^2=(\mfg^2\rho)^{-1}$ brings the metric (\ref{BDHM:tilde}) into the FG form (\ref{FG}), thereby justifying the application of holographic renormalization to the Euclidean BDHM background in the coordinate system \eqref{BDHM:tilde}. Under this transformation, the radial cutoff $\tir=\tir_\infty$ is related to the FG cutoff $\epsilon$ used in (\ref{ct-inf:log}) via $\tir_\infty^2=(\mfg^2\epsilon)^{-1}$. 

\subsection{Euclidean on-shell actions}\label{sec:OSA:on-shell}
Here we compute the regularized Euclidean on-shell action (\ref{I:reg}) for the BDHM background in the asymptotically non-rotating frame (\ref{BDHM:tilde}).

\bigskip
\noindent\textbf{Bulk action}
\medskip

\noindent We begin by evaluating the bulk on-shell action \eqref{S:E:onshell} for the Euclidean BDHM background~\eqref{BDHM:tilde}. As discussed in Section~\ref{sec:OSA:frame}, it is crucial to introduce a radial cutoff for the new radial coordinate $\tir$, not for $r$, to correctly capture the on-shell actions. However, performing the full 7d integral directly in the new coordinate system is technically involved. To proceed efficiently, we adopt the following strategy.
\begin{enumerate}
	\item Perform the trivial integrals over the four angular coordinates $(\tau,\phi_1,\phi_2,\phi_3)$. Recall that, after Wick rotation, the period of Euclidean time $\tau$ is related to the temperature as \cite{Bobev:2023bxl}
	\begin{align}
		(\tau,\phi_i)\sim(\tau+\beta,\phi_i-\ri\Omega_i\beta)\,, \qquad \beta=2\pi\fft{2\sqrt{\mS(r_+)}}{r_+^2(V'(r_+)-2mW'(r_+))}\,,\label{tau:beta}
	\end{align}
	which in turn is determined by the regularity condition of the near-horizon geometry ensuring the absence of conical singularities. The expressions for the angular velocities $\Omega_i$ are somewhat involved and can be found in \cite{Bobev:2023bxl}.
	
	\item Evaluate the radial integral over the range $\tir\in(\tir_+,\tir_\infty)$, where the lower bound $\tir_+$ denotes the horizon location in the new coordinate system. Using the relation (\ref{tir:ryz}), one can express $\tir_+$ in terms of $(r_+,y,z)$ as
	\begin{align}
		\tir_+=\tir_+(r_+,y,z)=\sqrt{\fft{(1+\mfg^2r_+^2)(1-\mfg^2y^2)(1-\mfg^2z^2)}{\mfg^2\Xi_1\Xi_2\Xi_3}-\fft{1}{\mfg^2}}\,.
	\end{align}

	\item Perform the remaining integrals over the coordinates $(y,z)$.
\end{enumerate}
With this procedure, the resulting bulk on-shell action \eqref{S:E:onshell} evaluated on the Euclidean BDHM background \eqref{BDHM:tilde} can be expressed in terms of the horizon radius $r_+$ in the original Boyer–Lindquist type coordinates. The result is
\begin{align}
	S_\text{bulk}\Big|_\text{(\ref{BDHM:tilde})}&=\fft{\beta\pi^2\mfg^2}{8G_N}\tir_\infty^6-\fft{3\beta\pi^2m(s_1^2+s_2^2)}{10G_N}(\mfg^2\tir_\infty^2+1)\sum_{i=1}^3\fft{\log\Xi_i}{(\Xi_i-\Xi_j)(\Xi_i-\Xi_k)}\nn\\
	&\quad-\fft{\beta\pi^2 m}{8G_N\Xi_1\Xi_2\Xi_3}\bigg(1+\fft{s_1^2+s_2^2}{2}(4-\Sigma_1)-\fft{(c_1-c_2)^2}{4}\Big(2\Sigma_2+8\Pi_1 \nn \\
	&\kern8em~ +\Sigma_1\prod_{i=1}^3(\Sigma_1-2a_i\mfg)\Big)\bigg)+\beta \mI\,,\label{BDHM:bulk}
\end{align}
where the symbol $\mI$ was defined in (4.29) of \cite{Bobev:2023bxl} and reads
\begin{align}
	\mI&=\fft{\pi^2}{16G_N\Xi_1\Xi_2\Xi_3r_+^2}\Bigg[(1-\mfg^2r_+^2){\textstyle\prod_{i=1}^{3}}(r_+^2+a_i^2)-2m(s_1^2+s_2^2)(\mfg^2r_+^4+a_1a_2a_3\mfg)\notag\\
	&\quad-\fft{2m^2s_1^2s_2^2}{\mfg^4\mS(r_+)}\Big({\textstyle\prod_{i=1}^{3}}(r_+^2+a_i^2)+2ms_1^2(r_+^2-a_1a_2a_3\mfg)\Big)\notag\\
	&\qquad\times\Big(\mfg^6r_+^6+(\Sigma_2+2\Pi_1)\mfg^4r_+^4-(2\Pi_1+\fft12\Pi_{22}-2\mfg^4ms_2^2)\mfg^2r_+^2+\Pi_1(2\mfg^4ms_2^2-\Pi_1)\Big)\notag\\
	&\quad-\fft{2m^2s_1^2s_2^2}{\mfg^4\mS(r_+)}\Big({\textstyle\prod_{i=1}^{3}}(r_+^2+a_i^2)+2ms_2^2(r_+^2-a_1a_2a_3\mfg)\Big)\notag\\
	&\qquad\times\Big(\mfg^6r_+^6+(\Sigma_2+2\Pi_1)\mfg^4r_+^4-(2\Pi_1+\fft12\Pi_{22}-2\mfg^4ms_1^2)\mfg^2r_+^2+\Pi_1(2\mfg^4ms_1^2-\Pi_1)\Big)\notag\\
	&\quad+\fft{m(c_1-c_2)^2}{\mfg^{14}\mS(r_+)}\prod_{i=1}^3\bigg(a_i\mfg+\fft{\Pi_1}{a_i\mfg}\bigg)\notag\\
	&\quad\times\bigg(-2\Xi_1\Xi_2\Xi_3\Big(\mfg^6r_+^6+\Sigma_2\mfg^4r_+^4+(\tfrac{1}{2}\Pi_{22}+2m \mfg^4(s_1^2+s_2^2))\mfg^2r_+^2+\Pi_2\notag\\
	&\qquad-2\mfg^4m(1+s_1^2+s_2^2)-\fft12(c_1-c_2)^2\mfg^4m(-2\Sigma_2-8\Pi_1+\Sigma_4-\Pi_{22})+(s_1^2+s_2^2)\mfg^4m\Sigma_2\Big)\notag\\
	&\qquad+\mfg^4m\Big(4+8\mfg^4ms_1^2s_2^2+(2-\Sigma_2-2\Pi_1)(2(s_1^2+s_2^2)-(c_1-c_2)^2(\Sigma_2+2\Pi_1))\Big)\notag\\
	&\qquad\times\fft{\mfg^8r_+^2{\textstyle\prod_{i=1}^{3}}(r_+^2+a_i^2)-\Xi_1\Xi_2\Xi_3}{1+\mfg^2r_+^2}-8\mfg^8m^2s_1^2s_2^2(2\mfg^6r_+^6+\Sigma_2\mfg^4r_+^4-\Pi_2)\bigg)\Bigg]\,.\label{mI}
\end{align}

Obtaining the result (\ref{BDHM:bulk}) is a far cry from merely using \texttt{Integrate} in \texttt{Mathematica}. Therefore, to facilitate the reproducibility of our results, we provide some further details:
\begin{itemize}
\item The $\star \mV$ term in (\ref{S:E:onshell}) is calculated following 1-2-3 above and it produces all of the divergent pieces in (\ref{BDHM:bulk}) together with a finite piece which contains the $\log\Xi_i$ terms.
\item The entire expression under the total derivative in (\ref{S:E:onshell}) is, by construction, regular. This means that, upon using Stoke's theorem, this term only receives contribution from evaluating $\qty{\dots}$ at the asymptotic cutoff radius $\tilde{r}_\infty$. Since the expression is finite, it can equally well be calculated using the $r$-coordinate (and send $r_\infty \rightarrow \infty$) or the $\tilde{r}$-coordinate (and send $\tilde{r}_\infty \rightarrow \infty$). Upon integrating, the $\big(\hA_{(1)}^1 \wedge F_{(2)}^2 + \hA_{(1)}^2 \wedge F_{(2)}^1\big) \wedge \hA_{(3)}$ term vanishes, while the $\sum_{I = 1}^{2}{} \hA_{(1)}^I \wedge X_I^{-2} \star F_{(2)}^I$ term yields a simple finite expression containing no logarithms. As an additional check of the imposed regularity, we verify that $\qty{\dots}$ evaluated on $r = r_+$ is identically zero.
\item To handle the $F_{(4)} \wedge \widehat{A}_{(3)}$ term, we find it convenient to convert back to the irregular (non-hatted) potentials
\begin{align}\label{eq:re-writing_F4wA3}
	F_{(4)} \wedge \widehat{A}_{(3)} &= F_{(4)} \wedge A_{(3)} \nn \\
	&+ \quad d {\qty{\frac{\ri}{4 \mfg} F_{(4)} \wedge d{\tau} \wedge \qty(\Phi_1 A_{(1)}^2 + \Phi_2 A_{(1)}^1) - \frac{1}{2 \mfg} F_{(4)} \wedge A_{(2)}} } \,.
\end{align}
Then the $F_{(4)} \wedge A_{(3)}$ term is easy to integrate, following 1-2-3 above, and produces a simple finite expressions containing new $\log$ terms, which are not of the form $\log\Xi_i$. The expression under the total derivative in \eqref{eq:re-writing_F4wA3} is not regular. This means that upon using Stoke's theorem we obtain a contribution from both asymptotic infinity and the horizon $r = r_+$. Term by term the contributions at infinity vanish identically. The contributions from $r = r_+$ yield cumbersome expressions involving various $\log$ terms.
\item Upon combining all of the above contributions, the $\log$ terms which are not of the form $\log\Xi_i$ cancel. To simplify the final answer and obtain~(\ref{BDHM:bulk}) one also needs to repeatedly replace various instances of $m^2$ (and higher even powers of $m$) with expressions involving only $m$ and $r_+$. This is achieved by using the definition of the horizon locus: $V(r_+)-2mW(r_+)=0$.
\end{itemize}

\bigskip
\noindent\textbf{Gibbons-Hawking-York term}
\medskip

\noindent Evaluating the GHY term (\ref{GHY}) on-shell in the asymptotically non-rotating frame of the BDHM background (\ref{BDHM:tilde}) presents some challenges, as the new radial coordinate $\tir$ contains cross terms with the $\tmu_i$ coordinates. This makes the boundary calculations involving the hypersurface at the radial cutoff $\tir=\tir_\infty$ more complicated. However, we found that a careful asymptotic analysis of the BDHM background (\ref{BDHM:tilde}) enables an analytic calculation of the on-shell GHY term. The key observation is that the BDHM background (\ref{BDHM:tilde}) can be expanded in the large $\tir$ limit as
\begin{subequations}
	\begin{align}
		ds^2&=g_{\mu\nu}dx^\mu dx^\nu=\overbrace{\bg_{\mu\nu}dx^\mu dx^\nu}^{=d\bs^2}+\gamma_{\mu\nu}dx^\mu dx^\nu\,,\\[0.5em]
		\gamma_{\mu\nu}&=\begin{cases}
			\tir^{-6}\textstyle{\sum_{n=0}^\infty \gamma_{\tir\tir}^{(n)}\tir^{-2n}} & (\mu\nu=\tir\tir) \\
			\tir^{-5}\textstyle{\sum_{n=0}^\infty \gamma_{\tir a}^{(n)}\tir^{-2n}} & (\mu\nu=\tir\tmu_a) \\
			\tir^{-2}\textstyle{\sum_{n=0}^\infty \gamma_{ij}^{(n)}\tir^{-2n}} & (\mu\nu=x^i x^j)\\
			0 & (\text{otherwise})
		\end{cases}\,,
	\end{align}\label{BDHM:asymp}%
\end{subequations}
where $d\bs^2$ stands for the global EAdS$_7$ metric (\ref{EAdS7}) and the perturbation coefficients $\gamma^{(n)}_{\mu\nu}$ can be extracted from the BDHM background (\ref{BDHM:tilde}) by substituting the large $\tir$ expansions~(\ref{tilde:expansion}). Here, the 7d and 6d coordinates are denoted as $\{x^\mu\}=\{\tau,\tir,\tmu_1,\tmu_2,\phi_i\}$ and $\{x^i\}=\{\tau,\tmu_1,\tmu_2,\phi_i\}$ respectively. Note that the coordinate $\tmu_3$ is replaced in terms of $\tmu_{1,2}$ by using the constraint (\ref{coord:change}). The asymptotic expansion of the BDHM background (\ref{BDHM:asymp}) induces the following 6d metric at the boundary $\tir=\tir_\infty$:
\begin{align}
	ds_6^2&=h_{ij}dy^idy^j=ds^2\Big|_{\tir=\tir_\infty}\nn\\
	&=\bh_{ij}dy^idy^j+\alpha_{ij}dy^idy^j\qquad\Big(\bh_{ij}=\bg_{ij},~\alpha_{ij}=\gamma_{ij}\Big)\,,\label{BDHM:bdry}
\end{align}
where the 6d boundary coordinates are identified as $y^i=x^i$.

\medskip

Using the asymptotic expansion of the BDHM background (\ref{BDHM:asymp}) and the induced boundary metric (\ref{BDHM:bdry}), we can determine the following outward unit normal vector and the corresponding extrinsic curvature with respect to the boundary $\tir=\tir_\infty$:
\begin{align}
	\ell_\mu&=\delta^{\tir}{}_\mu\,,\nn\\
	\quad n^\mu&=\fft{\ell_\mu}{\sqrt{\ell^\nu\ell_\nu}}=\sqrt{\bg^{\tir\tir}}\bigg[\delta^\mu{}_{\tir}\Big(1-\fft12\gamma^{\tir}{}_{\tir}\Big)-\delta^\mu{}_a \gamma^a{}_{\tir}+\mO(\tir_\infty^{-9})\bigg]\,,\nn\\
	\quad K&=\bK\bigg(1-\fft12\bg^{\tir\tir}\gamma_{\tir\tir}\bigg)+\fft{\sqrt{\bg^{\tir\tir}}}{2}\partial_{\tir}\alpha-\sqrt{\bg^{\tir\tir}}\fft{\partial_a(\sqrt{\bg}\bg^{ab}\gamma_{\tir b})}{\sqrt{\bg}}+\mO(\tir_\infty^{-8})\,.\label{BDHM:K}
\end{align}
Here, 7d and 6d coordinate indices are raised/lowered by the unperturbed metrics $\bg_{\mu\nu}$ and $\bh_{ij}$ respectively; for instance, $\alpha=\bh^{ij}\alpha_{ij}$. Substituting the extrinsic curvature (\ref{BDHM:K}) along with the boundary volume element
\begin{align}
	\sqrt{h}=\sqrt{\bh}\bigg[1+\fft12\alpha+\mO(\tir_\infty^{-8})\bigg]\,,\label{BDHM:bdry:vol}
\end{align}
into the GHY term (\ref{GHY}) and then evaluating the 6d boundary integral, we obtain its on-shell value 
\begin{align}
	S_\text{GHY}\Big|_\text{(\ref{BDHM:tilde})}&=-\fft{3\beta\pi^2\mfg^2}{4G_N}\tir_\infty^6-\fft{5\beta\pi^2}{8G_N}\tir_\infty^4+\fft{9\beta\pi^2m(s_1^2+s_2^2)}{5G_N}\bigg(\mfg^2\tir^2+\fft{14}{9}\bigg)\sum_{i=1}^3\fft{\log\Xi_i}{(\Xi_i-\Xi_j)(\Xi_i-\Xi_k)}\nn\\
	&\quad+\fft{3\beta\pi^2 m}{4G_N\Xi_1\Xi_2\Xi_3}\bigg(1+\fft{s_1^2+s_2^2}{2}(4-\Sigma_2)\nn\\
	&\kern8em~-\fft{(c_1-c_2)^2}{4}\Big(2\Sigma_2+8\Pi_1+\Sigma_1\prod_{i=1}^3(\Sigma_1-2a_i\mfg)\Big)\bigg)\,.\label{BDHM:GHY}
\end{align}
We discuss further details of the intermediate steps of this calculation in Appendix~\ref{app:bdry}.

\bigskip
\noindent\textbf{Counterterms}
\medskip

\noindent The counterterms can also be computed analytically using the asymptotic expansion of the BDHM background (\ref{BDHM:asymp}) and the induced boundary metric (\ref{BDHM:bdry}). The on-shell values of the first two counterterms, presented in~(\ref{ct-inf:0der}) and~(\ref{ct-inf:2der}), for the BDHM background~(\ref{BDHM:tilde}) can be computed using the boundary volume element~(\ref{BDHM:bdry:vol}) and the asymptotic expansion of the boundary Ricci scalar
\begin{align}
	\mR=\bmR+(-\alpha^{ij}\bmR_{ij}+\onabla_i\onabla_j\alpha^{ij}-\onabla_i\onabla^i\alpha)+\mO(\tir_\infty^{-8})\,.\label{BDHM:bdry:R}
\end{align}
On the other hand, the on-shell values of the four- or higher-derivative terms, such as~(\ref{ct-inf:4der}), (\ref{ct-inf:log}), and~(\ref{ct-fin}), for the BDHM background~(\ref{BDHM:tilde}) turn out to be identical to those for the global EAdS$_7$ background~(\ref{EAdS7}) up to $\mO(\tir_\infty^{-2})$ suppressed terms, due to the asymptotic behavior~(\ref{BDHM:asymp}). Consequently, calculations are considerably more straightforward in these cases. Furthermore, this suggests that we need not consider the various finite counterterms~(\ref{ct-fin}) individually, as they become linearly dependent on one another for the global EAdS$_7$ background~(\ref{EAdS7}). Therefore, below we only present the on-shell value for the first type of finite counterterm~(\ref{ct-fin:1}).

\medskip

We relegate the detailed calculations needed to implement the discussion above to Appendix~\ref{app:bdry}, and here we summarize the infinite and finite counterterms~(\ref{ct-inf}) and~(\ref{ct-fin}) evaluated for the BDHM background (\ref{BDHM:tilde}). The first two leading counterterms read ($i,j,k$ are all distinct)
\begin{subequations}\label{BDHM:ct}
\begin{align}
	S_\text{ct-0}\Big|_\text{(\ref{BDHM:tilde})}&=\fft{5\beta\pi^2\mfg^2}{8G_N}\tir_\infty^6+\fft{5\beta\pi^2}{16G_N}\tir_\infty^4-\fft{5\beta\pi^2}{64\mfg^2G_N}\tir_\infty^2+\fft{5\beta\pi^2}{128\mfg^4G_N}\nn\\
	&\quad-\fft{3\beta\pi^2m(s_1^2+s_2^2)}{2G_N}\Big(\mfg^2\tir_\infty^2+\fft{4}{3}\Big)\sum_{i=1}^3\fft{\log\Xi_i}{(\Xi_i-\Xi_j)(\Xi_i-\Xi_k)}\nn\\
	&\quad-\fft{5\beta\pi^2m}{8G_N\Xi_1\Xi_2\Xi_3}\bigg(1+\fft{s_1^2+s_2^2}{2}(4-\Sigma_2)\nn\\
	&\qquad-\fft{(c_1-c_2)^2}{4}\Big(2\Sigma_2+8\Pi_1+\Sigma_1\prod_{i=1}^3(\Sigma_1-2a_i\mfg)\Big)\bigg)\,,\\
	S_\text{ct-1}\Big|_\text{(\ref{BDHM:tilde})}&=\fft{5\beta\pi^2}{16G_N}\tir_\infty^4+\fft{5\beta\pi^2}{32\mfg^2G_N}\tir_\infty^2-\fft{5\beta\pi^2}{128\mfg^4G_N}\nn\\
	&\quad-\fft{\beta\pi^2m(s_1^2+s_2^2)}{2G_N}\sum_{i=1}^3\fft{\log\Xi_i}{(\Xi_i-\Xi_j)(\Xi_i-\Xi_k)}\,,
\end{align}
while the subleading counterterms that can be efficiently evaluated using the global EAdS$_7$ background are given by
\begin{align}
	S_\text{ct-2}\Big|_\text{(\ref{BDHM:tilde})}&=-\fft{5\beta\pi^2}{64\mfg^2G_N}\tir_\infty^2-\fft{5\beta\pi^2}{128\mfg^4G_N}\,,\\
	S_\text{ct-3}\Big|_\text{(\ref{BDHM:tilde})}&=0\,,\\
	S_\text{ct-fin-1}\Big|_\text{(\ref{BDHM:tilde})}&=\fft{1000\beta\pi^2 }{\mfg^4G_N}\,.
\end{align}
\end{subequations}
Note that the BDHM background of interest is free from holographic Weyl anomalies, indicated by the vanishing logarithmic counterterm.

\bigskip
\noindent\textbf{Regularized Euclidean on-shell action}
\medskip

\noindent Substituting the above Euclidean on-shell actions for the bulk term (\ref{BDHM:bulk}), the GHY term~(\ref{BDHM:GHY}), and the counterterms (\ref{BDHM:ct}) into the formula (\ref{I:reg}) and adding them all together, we obtain the regularized on-shell action of the BDHM background~(\ref{BDHM:tilde}) which reads:
\begin{align}
	I_\text{reg}\Big|_\text{(\ref{BDHM:tilde})}=\beta \mI-\fft{5\beta\pi^2}{128\mfg^4G_N}\big(1-25600\mtc_1\big)\,,\label{BDHM:Ireg}
\end{align}
where $\mI$ is presented in (\ref{mI}) and we have not included any finite counterterms except for the cubic Ricci scalar term~(\ref{ct-fin:1}) since, as discussed above, they are linearly dependent.

\medskip

We would like to highlight that, to the best of our knowledge, this represents the first rigorous calculation of the Euclidean on-shell action for electrically charged, rotating AdS$_7$ black holes in 7d maximal gauged supergravity, achieved through proper holographic renormalization. In previous studies, even for special cases with a reduced parameter space -- such as identical rotations \cite{Chong:2004dy}, identical charges \cite{Chow:2007ts}, and single rotation \cite{Wu:2011gp} -- the regularized on-shell action was computed indirectly, relying on the quantum statistical relation between the conserved charges of the black hole background and the corresponding chemical potentials, see also \cite{Cassani:2019mms,Kantor:2019lfo}. Hence, our calculation places a holographic comparison between the regularized on-shell action of the Euclidean BDHM background and the dual $S^1 \times S^5$ superconformal index (SCI) of the 6d $\mN=(2,0)$ theory on a more robust footing, without resorting to black hole thermodynamics. In what follows, we discuss the regularized on-shell action of the BDHM background (\ref{BDHM:Ireg}) in this context of holography, with particular emphasis on the choice of finite counterterms.

\subsection{Holography}\label{sec:OSA:holo}
The AdS/CFT correspondence relates the Euclidean M-theory path integral around the 11d supergravity background obtained by uplifting the supersymmetric limit of the BDHM solution via the consistent truncation of~\cite{Nastase:1999cb,Nastase:1999kf,Liu:1999ai} to the dual SCI defined as supersymmetric path integral of the $\mN=(2,0)$ $A_N$ SCFT on $S^1 \times S^5$~\cite{Choi:2018hmj,Hosseini:2018dob,Cassani:2019mms,Kantor:2019lfo,Nahmgoong:2019hko,Bobev:2015kza,Ohmori:2020wpk,Bobev:2023bxl}. In the limit where the AdS radius is much larger than the 11d Planck length, the M-theory path integral is approximated by the regularized on-shell action of the supersymmetric Euclidean BDHM background in 7d gauged supergravity. According to the AdS/CFT correspondence, this regularized on-shell action should agree with the large $N$ limit of the dual SCI. This relation was explicitly verified in \cite{Bobev:2023bxl} provided that
\begin{align}
	I_\text{reg}\Big|_\text{(\ref{BDHM:tilde})}=\beta \mI-\fft{5\beta\pi^2}{128\mfg^4G_N}\big(1-25600\mtc_1\big)\overset{!}{=}\beta \mI\,.\label{BDHM:Ireg:holo}
\end{align}

This holographic agreement between the two path integrals strongly motivates a particular choice of finite counterterms, i.e. setting $\mtc_1 = \fft{1}{25600}$. Alternative combinations of various finite counterterms presented in (\ref{ct-fin}) -- or possibly beyond -- are possible, and they will lead to different values of the coefficients $\mtc_x$. However, these counterterms are linearly dependent for backgrounds that approach the global EAdS$_7$ geometry (\ref{EAdS7}) in the asymptotic region, implying that such variations lead to equivalent results in the context of holographic renormalization. We thus adopt the choice $\mtc_1 = \fft{1}{25600}$ with all other coefficients vanishing, i.e. $\mtc_x = 0\,(x\neq1)$, and comment further on this holography-motivated choice of finite counterterms below. 

\medskip

Importantly, the first law of black hole thermodynamics for the BDHM solution is insensitive to the choice of finite counterterms. This is because finite counterterms contribute only constant shifts to the on-shell action, which are independent of the six parameters $(m,a_1,a_2,a_3,\delta_1,\delta_2)$ characterizing the black hole. This fact can be attributed to the choice of a proper coordinate system where the asymptotic behavior of the black hole matches the global EAdS$_7$ metric (\ref{EAdS7}) as discussed in~\cite{Gibbons:2005jd}. This fact implies that black hole thermodynamics, by itself, cannot justify a particular choice of finite counterterms.

\medskip

Another important point is that applying our choice of finite counterterms to the global EAdS$_7$ background, i.e. empty AdS$_7$, yields a vanishing regularized on-shell action. Therefore, it does not resolve the mismatch issue in the AdS$_7$/CFT$_6$ correspondence regarding the $\fft{81}{80}$ factor of discrepancy between the regularized on-shell action of global EAdS$_7$ and the dual field theory quantity, which is encoded in the $S^5$ free energy of 5d SYM through the perspective originally proposed in \cite{Douglas:2010iu,Lambert:2010iw}, as discussed in~\cite{Kallen:2012zn,Minahan:2013jwa,Bobev:2019bvq}. The standard local counterterms introduced in subsection~\ref{sec:OSA:action} are insufficient to resolve this mismatch, and one may instead consider giving up gauge invariance of the counterterms as suggested in~\cite{Bobev:2019bvq}. Thus, despite the progress we make here, a 
unified understanding of holographic renormalization, including finite counterterms, for all asymptotically locally EAdS$_7$ backgrounds remains a subject for future work.

\medskip

Lastly, the quantum statistical relation (QSR) \cite{Gibbons:1976ue} for the AdS$_7$ black hole of interest, namely
\begin{align}
	I_\text{reg}=-S+\beta\Bigg[E-\sum_{i=1}^3\Omega_iJ_i-\sum_{I=1}^2\Phi_IQ_I\Bigg]\,,
\end{align}
holds for the regularized on-shell action under the holography-motivated choice of finite counterterms, along with the black hole charges and potentials, as shown in \cite{Bobev:2023bxl}. In this verification, it is worth noting that the black hole energy $E$ was computed using the prescription of \cite{Ashtekar:1999jx}, which yields zero for global AdS$_7$ by construction. Hence, one may interpret our choice of finite counterterms as consistent with the prescription that sets the `vacuum (Casimir) energy' of global AdS$_7$ to zero. It should be noted that there is an alternative renormalization scheme in which the ``supersymmetric Casimir energy'' of empty AdS$_7$ is non-zero and should be given by the large $N$ limit of the $S^1 \times S^5$ free energy on ``the first sheet'', see \cite{Bobev:2015kza}. As discussed above the holographic implementation of this renormalization scheme is not entirely clear.\footnote{See \cite{BenettiGenolini:2016tsn,Papadimitriou:2019gel,Katsianis:2019hhg} for a discussion of the analogous problem in 5d supergravity and 4d $\mathcal{N}=1$ SCFTs.}

It is then natural to ask what happens if one changes the choice of finite counterterms. The QSR would no longer hold for the resulting regularized on-shell action, assuming the energy is still computed using the method of \cite{Ashtekar:1999jx}. To restore the QSR, one must evaluate the black hole energy using the counterterm method introduced in \cite{Balasubramanian:1999re}; more precisely, both the on-shell action and the energy must be computed using exactly the same set of infinite and finite counterterms. In this sense, the confirmation of the QSR in \cite{Bobev:2023bxl} was effectively achieved using the holography-motivated finite counterterm choice, where the vacuum energy is set to zero. A different choice of finite counterterms leads to a non-zero vacuum energy, as observed in \cite{Balasubramanian:1999re}. However, we emphasize that this vacuum energy in the AdS$_7$ context has not been fully matched with dual CFT computations \cite{Gibbons:2005jd}, while the duality between the BDHM on-shell action and the SCI that we have exhibited here strongly supports our holography-motivated choice of finite counterterms.

\section{AdS$_7$ black holes with generalized conformal boundaries}\label{sec:newBH}
In this section, inspired by the form of the BDHM solution presented in Section~\ref{sec:setup:sugra}, we construct a new class of AdS$_7$ black holes by generalizing the conformal boundary, focusing on the case with equal rotation parameters. 

\subsection{5d Sasaki-Einstein manifolds}\label{sec:newBH:SE5}
We start with a review of the class of 5d Sasaki-Einstein (SE) manifolds denoted by $L^{p,q,r}$~\cite{Cvetic:2005ft,Cvetic:2005vk}, which will be employed to generalize the AdS$_7$ BDHM black hole in Section~\ref{sec:newBH:newBH}.

\medskip

The local geometry of $L^{p,q,r}$ is described by the metric
\begin{subequations}
\begin{align}
	ds_{L^{p,q,r}}^2&=\tsigma^2+ds_{B_4}^2\,,\\
	\tsigma&=d\psi+(1-x/\mta)\sin^2\theta d\phi_1+(1-x/\mtb)\cos^2\theta d\phi_2\,,\\
	ds_{B_4}^2&=\fft{\rho^2}{4\Delta_x}dx^2+\fft{\rho^2}{\Delta_\theta}d\theta^2+\fft{\Delta_x}{\rho^2}\bigg(\fft{\sin^2\theta}{\mta}d\phi_1+\fft{\cos^2\theta}{\mtb}d\phi_2\bigg)^2\nn\\
	&\quad+\fft{\Delta_\theta\sin^2\theta\cos^2\theta}{\rho^2}\bigg(\fft{\mta-x}{\mta}d\phi_1-\fft{\mtb-x}{\mtb}d\phi_2\bigg)^2\,,
\end{align}\label{Lpqr}%
\end{subequations}
with the functions 
\begin{subequations}
\begin{align}
	\Delta_x&=x(\mta-x)(\mtb-x)-\mu\,,\\
	\Delta_\theta&=\mta\cos^2\theta+\mtb\sin^2\theta\,,\\
	\rho^2&=\Delta_\theta-x\,.
\end{align}
\end{subequations}
The local metric depends on three parameters $\{\mta, \mtb, \mu\}$. One of them can be set to unity by rescaling the coordinate $x$ and the remaining two parameters, leaving two independent real parameters that specify the metric. It is worth mentioning that the $L^{p,q,r}$ metric~(\ref{Lpqr}) is \emph{not} locally equivalent to the $S^5$ metric for generic $\mu\neq0$, as is evident from the coordinate-dependent curvature invariants~\cite{Cvetic:2005vk}.

\medskip

For global regularity, the metric must be positive definite, and the degeneration of the $\U(1)\times\U(1)\times\U(1)$ principal orbits in (\ref{Lpqr}) must be carefully analyzed to ensure the absence of conical singularities. Detailed treatments of these regularity conditions can be found in \cite{Cvetic:2005ft,Cvetic:2005vk,Butti:2005sw,Sfetsos:2005kd}. As a consequence of these regularity conditions, the ranges of the $x$ and $\theta$ coordinates are
\begin{align}
	x_1\leq x\leq x_2\,,\qquad 0\leq\theta\leq\fft\pi2\,,
\end{align}
where $x_{1,2}$ are the two smallest real roots of $\Delta_x$. The normalized Killing vectors generating the collapsing orbits at the four degeneration loci lead to the identifications 
\begin{equation}
\begin{alignedat}{5}
	\theta&=0&~&:&\quad&\partial_{\phi_1}&\quad&\to&\quad \phi_1&\sim\phi_1+2\pi\,,\\
	\theta&=\fft\pi2&~&:&\quad&\partial_{\phi_2}&\quad&\to&\quad \phi_2&\sim\phi_2+2\pi\,,\\
	x&=x_i&~&:&\quad&\ell_i=\mfc_i\partial_\psi+\mfa_i\partial_{\phi_1}+\mfb_i\partial_{\phi_2}&\quad&\to&\quad (\psi,\phi_1,\phi_2)&\sim(\psi,\phi_1,\phi_2)+2\pi(\mfc_i,\mfa_i,\mfb_i)\,,
\end{alignedat}\label{Lpqr:range}
\end{equation}
where $(\mfa_i,\mfb_i,\mfc_i)$ are defined by 
\begin{align}
	\mfa_i=\fft{\mta}{x_i-\mta}\mfc_i\,,\qquad\mfb_i=\fft{\mtb}{x_i-\mtb}\mfc_i\,,\qquad\mfc_i=-\fft{(\mta-x_i)(\mtb-x_i)}{\Delta_x'(x_i)}\,.
\end{align}
Global regularity further imposes a linear relation among the four Killing vectors presented in (\ref{Lpqr:range}) as
\begin{align}
	p\ell_1+q\ell_2+r\partial_{\phi_1}+(p+q-r)\partial_{\phi_2}=0\,,
\end{align}
where $\{p,q,r\}$ are coprime integers obeying
\begin{align}
	0<p\leq q\qquad \text{and}\qquad 0<r<p+q\,.\label{pqr}
\end{align}
This condition fixes the parameters $\{\mta,\mtb,\mu\}$ characterizing the metric in terms of $\{p,q,r\}$, though the explicit expressions require solving a quartic equation \cite{Butti:2005sw} and are omitted here. The volume of the resulting $L^{p,q,r}$ manifold can be computed using the metric and coordinate ranges discussed above and reads \cite{Cvetic:2005ft,Cvetic:2005vk,Martelli:2005wy}
\begin{align}
	\Vol[L^{p,q,r}]=\fft{\pi^3|\mfc_1|(x_2-x_1)(\mta+\mtb-x_1-x_2)}{\mta\mtb q}\,.\label{eq:volLpqr}
\end{align}

\medskip

As discussed in \cite{Cvetic:2005ft,Cvetic:2005vk,Butti:2005sw}, the $L^{p,q,r}$ family of manifolds has $S^2\times S^3$ topology for general values of the parameters $(p,q,r)$ and includes several previously known geometries as special cases. For instance, when $p+q=2r$, it reduces to the $Y^{\bp,\bq}$ family \cite{Gauntlett:2004yd,Martelli:2004wu} with the identification $Y^{\bp,\bq}=L^{\bp-\bq,\bp+\bq,\bp}$. The $p = q = r = 1$ case yields the homogeneous $T^{1,1}$ space \cite{Castellani:1983tb,Romans:1984an,Gubser:1998vd}. Although the $p = 0$ case falls outside the allowed range of (\ref{pqr}), taking the formal limit $p\to0$ produces the orbifold $S^5/\mathbb{Z}_q$.

\subsection{AdS$_7$ black holes with $\mathbb{R}\times L^{p,q,r}$ conformal boundary}\label{sec:newBH:newBH}
As a first step toward the construction of a new class of AdS$_7$ black holes with a generalized conformal boundary, we restrict our attention to the AdS$_7$ black hole with three identical rotation parameters and two independent electric charges, which, for an $\mathbb{R}\times S^5$ boundary, was first constructed and analyzed in \cite{Chong:2004dy,Cvetic:2005zi}. Rather than following the original conventions of \cite{Chong:2004dy,Cvetic:2005zi}, we can simply identify the three rotation parameters as $a_i=-a$ in the most general BDHM AdS$_7$ black hole solution summarized in Section~\ref{sec:setup:bdhm}, following Appendix B of \cite{Bobev:2023bxl}. The resulting background can be written more compactly as
\begin{subequations}
\begin{align}
	ds^2&=(H_1H_2)^{\fft15}\Bigg[-\fft{U}{f_1}dt^2+\fft{r^2(r^2+a^2)^2}{U}dr^2+\fft{r^2+a^2}{\Xi}ds^2_{\mathbb{CP}^2}\\
	&\kern6em+\fft{f_1}{(r^2+a^2)^2H_1H_2\Xi^2}\left(\sigma+\mfg dt-\fft{2f_2}{f_1}(1+a\mfg)dt\right)^2\Bigg]\,,\nn\\
	A^I_{(1)}&=\bigg(1-\fft{1}{H_I}\bigg)\fft{1}{\Xi s_I}\Big(\alpha_I(1+a\mfg)dt-a\alpha_{I+1}(\sigma+\mfg dt)\Big)\,,\\
	A_{(2)}&=\left(\fft{1}{H_1}+\fft{1}{H_2}\right)\fft{mas_1s_2}{(1+ag)(r^2+a^2)^2}dt\wedge\sigma\,,\\
	A_{(3)}&=\fft{mas_1s_2}{\Xi(1+a\mfg)(r^2+a^2)}(\sigma+\mfg dt)\wedge d\sigma\,,
\end{align}\label{BDHM:equal}%
\end{subequations}
where the parameters and functions are introduced as
\begin{subequations}
\begin{align}
	\Xi&= 1-a^2\mfg^2\,,\\
	s_I&=\sinh\delta_I\,,\qquad c_I=\cosh\delta_I\,,\qquad(\delta_{I+2}=\delta_I)\\
	\alpha_I&=c_I-\fft12(1-(1-a\mfg)^2)(c_I-c_{I+1})=\alpha_{I+2}\,,\\
	H_I(r)&=1+\fft{2ms_I^2}{(r^2+a^2)^2}\,,\\
	f_1(r)&=(1-a^2\mfg^2)H_1H_2(r^2+a^2)^3-\fft{4(1-a\mfg)^2m^2a^2s_1^2s_2^2}{(r^2+a^2)^2}\\
	&\quad+\fft12ma^2\left(4(1-a\mfg)^2+2(1-(1-a\mfg)^4)c_1c_2+(1-(1-a\mfg)^2)^2(c_1^2+c_2^2)\right)\,,\nn\\
	f_2(r)&=\fft12\mfg(1-a\mfg)H_1H_2(r^2+a^2)^3 \\
	&\quad+\fft14ma\left(2(1+(1-a\mfg)^4)c_1c_2+(1-(1-a\mfg)^4)(c_1^2+c_2^2)\right)\,,\nn\\
	U(r)&=\mfg^2H_1H_2(r^2+a^2)^4+(1-a^2\mfg^2)(r^2+a^2)^3+\fft12ma^2\left(4(1-a\mfg)^2 \right. \\
	& \left. \quad + 2(1-(1-a\mfg)^4)c_1c_2 +(1-(1-a\mfg)^2)^2(c_1^2+c_2^2)\right)\nn\\
	&\quad-\fft12m(r^2+a^2)\left(4(1-a^2\mfg^2)+2a^2\mfg^2(6-8a\mfg+3a^2\mfg^2)c_1c_2 \right. \nn\\
	& \quad \left. -a^2\mfg^2(2-a\mfg)(2-3a\mfg)(c_1^2+c_2^2)\right)\,.\nn
\end{align}
\end{subequations}
Note that we have cyclically identified the electric charge parameters for notational convenience. In the presentation above, the original $S^5$ metric $ds_{S^5}^2$ given in (\ref{EAdS7}) is rewritten as a $\U(1)$ fibration over the Fubini-Study metric on $\mathbb{CP}^2$ as follows
\begin{subequations}
\begin{align}
	ds_{S^5}^2&=\sigma^2+ds^2_{\mathbb{CP}^2}\,,\\
	\sigma&=d\psi+\fft12\sin^2\xi\,\sigma_3\,,\\
	ds^2_{\mathbb{CP}^2}&=d\xi^2+\fft14\sin^2\xi(\sigma_1^2+\sigma_2^2)+\fft14\sin^2\xi\cos^2\xi \sigma_3^2\,,
\end{align}\label{S5}%
\end{subequations}
where the $\SU(2)$ left-invariant 1-forms are given by
\begin{equation}
	\begin{split}
		\sigma_1&=\cos\varphi_3 d\varphi_1+\sin\varphi_1\sin\varphi_3 d\varphi_2\,,\\
		\sigma_2&=\sin\varphi_3 d\varphi_1-\sin\varphi_1\cos\varphi_3 d\varphi_2\,,\\
		\sigma_3&=d\varphi_3+\cos\varphi_1 d\varphi_2\,.
	\end{split}\label{SU(2):LI}
\end{equation}
For completeness, we reiterate that the necessary for regularity gauge shifts, discussed in Section \ref{sec:setup:gg}, are
\begin{align}
  \begin{aligned}
  A_{(1)}^I \rightarrow {}& \widehat{A}^I_{(1)} = A_{(1)}^I - \Phi_I dt \,, \\
  A_{(3)} \rightarrow {}& \widehat{A}_{(3)} = A_{(3)} + \frac{1}{4 \mfg} dt \wedge (\Phi_1 F_{(2)}^2 + \Phi_2 F_{(2)}^1) - \frac{1}{2 \mfg} d A_{(2)} \,. 
  \end{aligned}
\end{align}

\medskip

The key observations underlying the generalization of the AdS$_7$ black hole background with equal rotation parameters (\ref{BDHM:equal}) are as follows.
\begin{itemize}
	\item In the original solution (\ref{BDHM:equal}), the $S^5$ metric (\ref{S5}) enters the geometry only through the 1-form $\sigma$ and the 4d base space metric $ds^2_{\mathbb{CP}^2}$.
	
	\item The $L^{p,q,r}$ family of SE$_5$ manifolds admits a metric written as a $\U(1)$ fibration over a 4d K\"ahler base as presented in (\ref{Lpqr}).
\end{itemize}
Motivated by these observations, we propose the following 7d supergravity backgrounds which form a large class of new AdS$_7$ black holes with conformal boundary $\mathbb{R}\times L^{p,q,r}$: 
\begin{subequations}
	\begin{align}
		ds^2&=(H_1H_2)^{\fft15}\Bigg[-\fft{U}{f_1}dt^2+\fft{r^2(r^2+a^2)^2}{U}dr^2+\fft{r^2+a^2}{\Xi}ds^2_{B_4}\\
		&\kern6em+\fft{f_1}{(r^2+a^2)^2H_1H_2\Xi^2}\left(\tsigma+\mfg dt-\fft{2f_2}{f_1}(1+a\mfg)dt\right)^2\Bigg]\,,\nn\\
		A^I_{(1)}&=\bigg(1-\fft{1}{H_I}\bigg)\fft{1}{\Xi s_I}\Big(\alpha_I(1+a\mfg)dt-a\alpha_{I+1}(\tsigma+\mfg dt)\Big)\,,\\
		A_{(2)}&=\left(\fft{1}{H_1}+\fft{1}{H_2}\right)\fft{mas_1s_2}{(1+a\mfg)(r^2+a^2)^2}dt\wedge\tsigma\,,\\
		A_{(3)}&=\fft{mas_1s_2}{\Xi(1+a\mfg)(r^2+a^2)}(\tsigma+\mfg dt)\wedge d\tsigma\,,
	\end{align}\label{AdS7BH:Lpqr}%
\end{subequations}
where $\tsigma$ and $ds^2_{B_4}$ are presented in (\ref{Lpqr}). Note that the Ansatz (\ref{AdS7BH:Lpqr}) is obtained from the known black hole background (\ref{BDHM:equal}) through the straightforward substitutions
\begin{align}
	\sigma~\to~\tsigma\qquad\text{and}\qquad ds^2_{\mathbb{CP}^2}~\to~ds^2_{B_4}\,.\label{substitution}
\end{align}
We have confirmed that the background (\ref{AdS7BH:Lpqr}) is indeed a solution of the $\U(1)\times\U(1)$ truncation of 7d maximal gauged supergravity, satisfying the equations of motion~(\ref{eom}) along with the self-duality constraint~(\ref{self-dual}). In particular, the equations of motion have been verified for general values of the parameters with $\mu\neq 0$, demonstrating that the local solution (\ref{AdS7BH:Lpqr}) constitutes a non-trivial extension of the known solution (\ref{BDHM:equal}) already at the local level. 

\medskip

To obtain a globally well-defined geometry, we impose the regularity constraints associated with the collapsing $\U(1)$ orbits in the background (\ref{AdS7BH:Lpqr}) so as to avoid conical singularities. The resulting conditions reproduce the global properties of the $L^{p,q,r}$ space summarized in Section~\ref{sec:newBH:SE5} and ensure that the horizon of the black hole solution is the same as the $L^{p,q,r}$ Sasaki-Einstein manifold. Regularity of the 7d solution in Euclidean signature requires that the thermal circle (with $t=-\ri\tau$) and the angular coordinate $\psi$ have the following periodicities
\begin{align}
	(\tau,\psi)\sim(\tau+\beta,\psi+\ri\Omega\beta)\,,\label{thermal:cycle}
\end{align}
where the inverse temperature $\beta$ and the angular velocity $\Omega$ are presented in (\ref{T}) and (\ref{Omega}) respectively. These global properties complete the construction of the family of AdS$_7$ black hole backgrounds with generalized conformal boundary $\mathbb{R}\times L^{p,q,r}$, complementing the local description (\ref{AdS7BH:Lpqr}) and ensuring the absence of conical singularities. Note that this new class of AdS$_7$ black holes is asymptotically \emph{locally} AdS$_7$. Their conformal boundary, as well as the horizon sections, are described by $\mathbb{R} \times L^{p,q,r}$, meaning that the asymptotic region does not match global EAdS$_7$ with $S^5$ slices.

\medskip

The Lorentzian backgrounds defined above may suffer from naked closed-timelike-curves (CTCs). To analyze whether such causal pathologies are present one can proceed exactly as in the $\mathbb{R}\times S^5$ case detailed in \cite{Cvetic:2005zi}, since the analysis is insensitive to the substitution~(\ref{substitution}). Hence, as in the special case with the $\mathbb{R}\times S^5$ conformal boundary, the background (\ref{AdS7BH:Lpqr}) generically exhibits naked CTCs, although these can be avoided for certain  special cases as discussed in \cite{Cvetic:2005zi} -- see also \cite{Chow:2007ts} for a similar observation in backgrounds with independent rotation parameters. In this sense, the large class of AdS$_7$ black holes with $\mathbb{R}\times L^{p,q,r}$ conformal boundary constructed above is as globally well-defined as the solutions with $S^5$ horizon and $\mathbb{R}\times S^5$ boundary studied in~\cite{Chong:2004dy,Cvetic:2005zi}.

\subsection{Thermodynamics and supersymmetric limits}\label{sec:newBH:ThDsusy}
We now proceed to examine the thermodynamics of the new class of AdS$_7$ black holes constructed in Section~\ref{sec:newBH:newBH}, along with their supersymmetric limit. As the analysis closely parallels the BDHM black hole case discussed in \cite{Bobev:2023bxl}, we omit some intermediate steps and focus on presenting the final expressions.

\subsubsection{Thermodynamics}\label{sec:newBH:ThDsusy:ThD}
We begin with the extensive thermodynamic quantities. The Bekenstein-Hawking entropy of the AdS$_7$ black hole (\ref{AdS7BH:Lpqr}) is
\begin{align}
	S=\fft{\Vol[L^{p,q,r}](r_+^2+a^2)\sqrt{f_1(r_+)}}{4G_N\Xi^3}\,,
\end{align}
where $r_+$ denotes the horizon radius given by the largest positive root of the radial function $U(r)$. The angular momentum is obtained from the Komar integral evaluated at spatial infinity and reads
\begin{align}
	J&=-\fft{1}{16\pi G_N}\int_{L^{p,q,r}}\star dK \label{J}\\[0.5em]
	&=\frac{a m \Vol[L^{p,q,r}]}{4 \pi  G_N \Xi ^4} \left(\alpha _1 \alpha _2 (1+ a\mfg)-\fft12 a\mfg\left(\alpha_1^2+\alpha_2^2\right)+\frac{1}{2} a\mfg (1-a\mfg)^2 \left(s_1^2 +s_2^2\right)\right)\,,\nn
\end{align}
where the 1-form $K=K_\mu dx^\mu$ is associated with the angular Killing vector $K^\mu\partial_\mu=\fft13\partial_\psi$.\footnote{The angular Killing vector is normalized in accordance with the coordinate transformation described in Appendix B of \cite{Bobev:2023bxl}, which gives $\partial_\psi = -\!\sum_{i=1}^3 \partial_{\phi_i}$, where the $\phi_i$ are the angular coordinates associated with the three independent rotation parameters $a_i$ describing the BDHM black hole. With this normalization, the angular momenta $J_i$ of \cite{Bobev:2023bxl} reduce to $J$ given in (\ref{J}) under the equal-rotation limit, up to the replacement of the 5d volume factor $\Vol[S^5]$ by its generalization $\Vol[L^{p,q,r}]$. \label{footnote:J}} The two electric charges are computed analogously as
\begin{align}
	Q_I&=-\fft{1}{16\pi G_N}\int_{L^{p,q,r}}\left( X^{-2}_I \star F^I_{(2)} - F^I_{(2)} \wedge A_{(3)} \right)\\[0.5em]
	&=\frac{m s_I \Vol[L^{p,q,r}]}{4 \pi G_{N} \Xi^3}\left(2 c_I - (c_I-c_{I+1}) a^2 \mfg^2 \left(3-2 a\mfg\right)\right)\,.\nn
\end{align}
For the calculation of the energy (or mass) of the AdS$_7$ black hole, we adopt the prescription of \cite{Ashtekar:1999jx} as implemented in \cite{Bobev:2023bxl}, and find
\begin{align}
	E&=\frac{1}{32 \pi \mfg^3} \int_{\Sigma} \mathrm{d} \widetilde{\Sigma}_\mu \, \widetilde{\Omega}^{-4} \widetilde{n}^\rho \widetilde{n}^\sigma \widetilde{C}^\mu{}_{\rho \nu \sigma} \xi^\nu\\[0.5em]
	&=\frac{m \Vol[L^{p,q,r}]}{16 \pi  G_N \Xi ^4}\bigg[(\alpha _1^2+\alpha _2^2) \left(5+10 a \mfg+11 a^2 \mfg^2\right) -20 a \alpha _1 \alpha _2 \mfg (1+a \mfg) \nn \\ 
	&\kern7em~ +3 \left(1+2a\mfg-a^2 \mfg^2\right) (1-a \mfg)^2 \left(s_1^2+s_2^2\right)\bigg]\,.\nn
\end{align}
Here, we introduce the conformally rescaled metric $\widetilde{g}_{\mu\nu} = \widetilde{\Omega}^{2}g_{\mu\nu}$ with $\widetilde{\Omega}=(\mfg r)^{-1}$ and $\widetilde{C}^{\mu}{}_{\nu\rho\sigma}$ denotes the Weyl tensor. The integration measure $d\widetilde{\Sigma}_{\mu}$ corresponds to the area element of the $L^{p,q,r}$ section at the conformal boundary. The timelike Killing vector $\xi$ and the normal vector $\widetilde{n}$ are specified by their components $\xi^\mu=\delta^\mu{}_t$ and $\widetilde{n}_{\mu} = \partial_{\mu}\widetilde{\Omega}$, respectively.

\medskip

We now turn to the intensive thermodynamic quantities. The Hawking temperature of the AdS$_7$ black hole (\ref{AdS7BH:Lpqr}) is determined by requiring the absence of conical singularity at the horizon in Euclidean signature
\begin{align}
	T=\beta^{-1}=\fft{U'\left(r_{+}\right)}{4 \pi  r_{+} \left(r_{+}^2+a^2\right) \sqrt{f_1\left(r_{+}\right)}}\,,\label{T}
\end{align}
where the inverse temperature $\beta$ fixes the range of Euclidean time coordinate $\tau=\ri t$, see~(\ref{thermal:cycle}). The angular velocity $\Omega$ can be found by specifying the Killing vector\footnote{The angular velocity is introduced in (\ref{ell}) following the convention specified in Footnote~\ref{footnote:J}, ensuring that the angular velocities $\Omega_i$ of~\cite{Bobev:2023bxl} reduce to $\Omega$ in~(\ref{Omega}) under the equal-rotation limit.}
\begin{align}
	\ell=\partial_t-\Omega\partial_\psi\,, \label{ell}
\end{align}
that generates the null horizon at $r=r_+$, which yields
\begin{align}
	\Omega=\mfg-\fft{2(1+a\mfg)f_2(r_+)}{f_1(r_+)}\,.\label{Omega}
\end{align}
The electric potentials, defined relative to the asymptotic region, are
\begin{align}
	\Phi_I=\ell^\mu A^I_{(1)\mu}\Big|_{r=r_+}-\ell^\mu A^I_{(1)\mu}\Big|_{r\to\infty}=\frac{2 m s_I \left(\alpha _I (1+a\mfg)-\alpha_{I+1}a(\mfg-\Omega )\right)}{\Xi (r_+^2+a^2)^2H_I}\,.
\end{align}

\medskip

All of the above expressions can be recovered directly from the BDHM black hole results \cite{Bobev:2023bxl} by taking the equal-rotation limit $a_i=-a$ and, for the extensive quantities, replacing the 5d volume factor $\Vol[S^5]=\pi^3$ with the $L^{p,q,r}$ volume $\Vol[L^{p,q,r}]$ in~\eqref{eq:volLpqr}. The first law of thermodynamics therefore holds automatically for these generalized AdS$_7$ black holes with $\mathbb{R}\times L^{p,q,r}$ conformal boundary (\ref{AdS7BH:Lpqr}) and reads
\begin{align}
	dE=TdS+3\Omega dJ+\sum_{I=1}^2\Phi_IdQ_I\,,
\end{align}
where the variations are taken with respect to the four parameters $(m,a,\delta_i)$ specifying the black hole.

\medskip

The regularized Euclidean on-shell action of the generalized AdS$_7$ black hole (\ref{AdS7BH:Lpqr}) can be computed via holographic renormalization following the procedure outlined in Section~\ref{sec:OSA}. Its explicit form can be obtained from the BDHM black hole result in \eqref{mI} and \eqref{BDHM:Ireg:holo} after setting $a_i=-a$ and replacing $\Vol[S^5]=\pi^3$ by $\Vol[L^{p,q,r}]$. Using the explicit form of the on-shell action and the black hole thermodynamic quantities we confirmed that the quantum statistical relation
\begin{align}
	I_\text{reg}\Big|_\text{(\ref{AdS7BH:Lpqr})}=-S+\beta\Bigg[E-3\Omega J-\sum_{I=1}^2\Phi_IQ_I\Bigg]\label{Ireg:Lpqr}
\end{align}
is indeed satisfied.

\subsubsection{Supersymmetry}\label{sec:newBH:ThDsusy:susy}
Following the supersymmetry analysis of \cite{Cvetic:2005zi}, as applied in \cite{Chow:2007ts,Cassani:2019mms,Kantor:2019lfo,Bobev:2023bxl}, we restrict our attention to the supersymmetric limit preserving two real supercharges, imposed by the following constraint on the black hole parameters:
\begin{align}
	-3a\mfg=\fft{2}{1-e^{\delta_1+\delta_2}}\,.\label{susy:constraint}
\end{align}
This condition arises from the BPS relation among the charges $(E,J,Q_I)$, whose expressions in the generalized AdS$_7$ $L^{p,q,r}$ black holes~(\ref{AdS7BH:Lpqr}) differ from those of the $S^5$ case only by an overall 5d volume factor. We therefore expect the same supersymmetry constraint (\ref{susy:constraint}) to apply more generally to AdS$_7$ black holes with $\mathbb{R}\times L^{p,q,r}$ conformal boundary. 

\medskip

Imposing the supersymmetric constraint (\ref{susy:constraint}) on the thermodynamic quantities presented in Section~\ref{sec:newBH:ThDsusy:ThD}, we recover the following relations among charges and chemical potentials
\begin{align}\label{eq:BPSconstLpqr}
	E+3\mfg J-Q_1-Q_2&=0\,,\\
	\beta\Big(\mfg-3\Omega-2\mfg\Phi_1-2\mfg\Phi_2\Big)&=\pm 2\pi\ri\,.
\end{align}
These relations also follow directly from the BDHM results \cite{Bobev:2023bxl} in the equal-rotation limit. For a detailed account of complexified chemical potentials in Euclidean signature, and the role of extremality for obtaining well-defined Lorentzian BPS AdS$_7$ black holes, we refer the reader to \cite{Bobev:2023bxl}.

\medskip

Finally, the regularized Euclidean on-shell action (\ref{Ireg:Lpqr}) takes the remarkably simple form in the supersymmetric limit (\ref{susy:constraint}) and reads
\begin{align}
	I_\text{reg}\Big|_\text{(\ref{AdS7BH:Lpqr})}\overset{\text{(\ref{susy:constraint})}}{=}\fft{\Vol[L^{p,q,r}]}{8\pi G_N\mfg}\fft{\varphi_1^2\varphi_2^2}{\omega^3}\,,\label{Ireg:Lpqr:susy}
\end{align}
where we have defined \cite{Bobev:2023bxl}
\begin{align}
	\varphi_I=\beta(\Phi_I-1)\qquad\text{and}\qquad \omega=\beta(\Omega+\mfg)\,.
\end{align}
The supersymmetric regularized Euclidean on-shell action (\ref{Ireg:Lpqr:susy}) is expected to capture, in the semi-classical limit, the supersymmetric path integral of the dual 6d $\mN=(2,0)$ theory over $S^1\times L^{p,q,r}$ on the `2nd sheet'. This result generalizes prior holographic comparisons for the 6d SCFT on $S^1\times S^5$~\cite{Hosseini:2018dob,Choi:2018hmj,Cassani:2019mms,Kantor:2019lfo,Bobev:2023bxl}. We now proceed to elaborate on this holographic interpretation.

\subsection{Holography}\label{sec:newBH:holo}

The supergravity background presented above is asymptotically locally AdS$_7$ with an $S^1 \times L^{p,q,r}$ boundary. It can be uplifted to 11d supergravity on $S^4$ and should therefore represent the holographic dual description of the 6d $\mathcal{N}=(2,0)$ $A_N$ SCFT placed on $S^1 \times L^{p,q,r}$. Confirming this holographic duality for the general non-supersymmetric supergravity background is currently not possible due to the lack of appropriate calculational tools on the QFT side. It is however reasonable to expect that some progress can be made in the supersymmetric limit of this setup.

Based on the $S^1 \times S^5$ results in \cite{Bobev:2015kza,Nahmgoong:2019hko}, their holographic agreement with the 7d BDHM supergravity solution, and the result~\eqref{Ireg:Lpqr:susy} above, we propose the following conjecture for the leading order large $N$, $S^1 \times L^{p,q,r}$ supersymmetric partition function of the $A_N$  6d $\mathcal{N}=(2,0)$ theory
\begin{equation}\label{eq:S1Lpqrconj}
-\log Z_{S^1\times L^{p,q,r}} = \frac{N^3}{24} \frac{\Vol[L^{p,q,r}]}{\pi^3} \frac{\Delta_1^2\Delta_2^2}{\hat{\omega}_1\hat{\omega}_2\hat{\omega}_3}\,,
\end{equation}
where we have used that $\Vol[S^5]=\pi^3$. The supersymmetry constraint between the R-symmetry and angular momentum fugacities reads
\begin{equation}
\Delta_1+\Delta_2 - \hat{\omega}_1 - \hat{\omega}_2 - \hat{\omega}_3 = 2\pi n \ri\,, \qquad n=0,\pm1\,.
\end{equation}
In our nomenclature, following~\cite{Cassani:2021fyv}, we refer to the choice $n=0$ as ``the first sheet'', while $n=\pm 1$ is called ``the second sheet''.

As a first consistency check of this proposal for the large $N$ SCFT partition function we note that it agrees with the supergravity result in~\eqref{Ireg:Lpqr:susy} in the limit of equal angular fugacities $\omega_i=\omega$ and on the ``second sheet'' $n=\pm1$. To show this we need the relation between the QFT and supergravity fugacities derived in \cite{Bobev:2023bxl} and given by 
\begin{equation}
\Delta_{I} = -2\mfg \beta (\Phi_I-1) = -2\mfg \varphi_I\,, \qquad \hat{\omega}_i = \beta(\Omega_i+\mfg) = \omega_i\,,
\end{equation}
as well as the holographic dictionary between the number of M5-branes and the gravitational parameters to leading order in the large $N$ limit
\begin{equation}
N^3 = \frac{3 \pi^2}{16 G_N \mfg^5}\,.
\end{equation}
Moreover, by design, the expression in \eqref{eq:S1Lpqrconj} agrees with the result for the $S^1\times S^5$ supersymmetric partition function in \cite{Bobev:2015kza,Nahmgoong:2019hko} which was confirmed also holographically in Section~\ref{sec:OSA} above as well as in~\cite{Bobev:2023bxl}.

The partition function of 6d SCFTs on $S^1\times Y^{p,q}$ and $S^1\times L^{p,q,r}$ was studied using supersymmetric localization in \cite{Qiu:2013pta} and \cite{Qiu:2014oqa}, respectively (see also \cite{Ruggeri:2025zon}). In the large $N$ limit the results in \cite{Qiu:2013pta,Qiu:2014oqa} agree with the conjecture in~\eqref{eq:S1Lpqrconj}, see Section 7 of \cite{Qiu:2013pta} and Section 5 of \cite{Qiu:2014oqa}. Importantly, in~\cite{Qiu:2013pta,Qiu:2014oqa} the large $N$ analysis was restricted to the case of equal electric and angular fugacities, i.e. $\Delta_I=\Delta$ and $\hat{\omega}_i=\hat{\omega}$, and to the first sheet $n=0$. While these QFT calculations support the conjecture in~\eqref{eq:S1Lpqrconj} they also underscore the open problem of constructing the holographic dual of the 6d SCFT on $S^1\times L^{p,q,r}$ on the first sheet, i.e. for $n=0$. For the theory on $S^1\times S^5$ this holographic dual is given simply by the empty AdS$_7$ solution in global coordinates. The analogous supergravity background with $S^1\times L^{p,q,r}$ boundary is however not smooth in the bulk and it is therefore unclear whether it is the proper description of the 6d SCFT on $S^1\times L^{p,q,r}$ with $n=0$.

Despite the strong evidence summarized above, it is important to put the conjecture in~\eqref{eq:S1Lpqrconj} on a more solid footing. To this end one should generalize the 6d analysis of \cite{Nahmgoong:2019hko} to $S^1\times L^{p,q,r}$ or work towards a 6d generalization of the $S^1\times S^3$ EFT for 4d $\mathcal{N}=1$ SCFTs studied in \cite{Cassani:2021fyv,ArabiArdehali:2021nsx,Ardehali:2021irq}. In both of these cases, the end result will be a check of~\eqref{eq:S1Lpqrconj} in the Cardy limit where the length of the $S^1$ is much smaller than the volume of $L^{p,q,r}$. It will be very interesting to pursue this analysis and derive~\eqref{eq:S1Lpqrconj} more rigorously.

\section{AdS$_5$ black holes with generalized conformal boundaries}\label{sec:newBH-5}
In this section, we briefly comment on the generalization of electrically charged rotating AdS$_5$ black holes in 5d gauged supergravity, motivated by the extension of AdS$_7$ black holes examined in the previous section.

\medskip

We focus on the 5d gauged supergravity coupled to two vector multiplets, referred to as the gauged STU model. In this theory, the most general black hole solution with two rotation parameters and three electric charge parameters was constructed in \cite{Wu:2011gq}. For the purpose of generalization here, we focus on a special case where the rotation parameters are identical, which was first found in \cite{Cvetic:2004ny} and further studied in \cite{Cvetic:2005zi,Cassani:2019mms,Kantor:2019lfo}. Following the conventions of \cite{Cvetic:2005zi,Cassani:2019mms}, the solution reads
\begin{align}
	ds^2&=(H_1H_2H_3)^\fft13\bigg[-\fft{r^2Y}{f_1}dt^2+\fft{r^4}{Y}dr^2+\fft14r^2(\sigma_1^2+\sigma_2^2)+\fft{f_1}{4r^4H_1H_2H_3}(\sigma_3-\fft{2f_2}{f_1}dt)^2\bigg]\,,\nn\\
	A_{(1)}^I&=\fft{2m}{r^2H_I}\bigg(s_Ic_Idt+\fft12a(c_Is_Js_K-s_Ic_Jc_K)\sigma_3\bigg)\,,\label{AdS5BH}\\
	X_I&=\fft{(H_1H_2H_3)^\fft13}{H_I}\,,\nn
\end{align}%
where we have employed the $\SU(2)$ left-invariant 1-forms (\ref{SU(2):LI}), and the various functions are defined as
\begin{subequations}
\begin{align}
	s_I&=\sinh\delta_I\,,\qquad c_I=\cosh\delta_I\,,\\
	H_I(r)&=1+\fft{2ms_I^2}{r^2}\,,\\
	f_1(r)&=r^6H_1H_2H_3+2ma^2r^2+4m^2a^2\big[2(c_1c_2c_3-s_1s_2s_3)s_1s_2s_3-{\scriptstyle\sum_{I<J}^3}s_I^2s_J^2\big]\,,\\
	f_2(r)&=2ma(c_1c_2c_3-s_1s_2s_3)r^2+4m^2as_1s_2s_3\,,\\
	f_3(r)&=2ma^2(1+g^2r^2)+4g^2m^2a^2\big[2(c_1c_2c_3-s_1s_2s_3)s_1s_2s_3-{\scriptstyle\sum_{I<J}^3}s_I^2s_J^2\big]\,,\\
	Y(r)&=f_3+g^2r^6H_1H_2H_3+r^4-2mr^2\,.
\end{align}
\end{subequations}
The angular coordinates describing the $\SU(2)$ left-invariant 1-forms (\ref{SU(2):LI}) have the following ranges and periodicities:
\begin{align}
	0\leq\varphi_1\leq\pi\,,\qquad (\varphi_2,\varphi_3)\sim(\varphi_2,\varphi_3)+2\pi(\pm 1,1)\,,\qquad \varphi_3\sim\varphi_3+4\pi\,.\label{L11}
\end{align}
Therefore, the AdS$_5$ black hole (\ref{AdS5BH}) has a conformal boundary $\mathbb{R} \times S^3$ in the asymptotic region ($r \to \infty$), with the unit radius $S^3$ metric
\begin{align}
	ds^2_{S^3}&=\fft14\sum_{i=1}^3\sigma_i^2=dz_1d\bar{z}_1+dz_2d\bar{z}_2\qquad\Big(z_1=e^{\ri\fft{\varphi_3+\varphi_2}{2}}\cos\fft{\varphi_1}{2},~z_2=e^{\ri\fft{\varphi_3-\varphi_2}{2}}\sin\fft{\varphi_1}{2}\Big)\nn\\
	&=\fft14\Big[(d\varphi_3+\cos\varphi_1d\varphi_2)^2+d\varphi_1^2+\sin^2\varphi_1d\varphi_2^2\Big]\,.\label{S3}
\end{align}

Motivated by the generalization of the AdS$_7$ black holes discussed in Section~\ref{sec:newBH}, we now extend the AdS$_5$ black hole background (\ref{AdS5BH}) by modifying its conformal boundary from $\mathbb{R} \times S^3$ to $\mathbb{R} \times L(p, q)$ with the lens space $L(p,q)$. This is achieved by maintaining the same local metric (\ref{AdS5BH}), but altering the periodic identification of the angular coordinates associated with the $\SU(2)$ left invariant 1-forms. Specifically, we replace the last periodicity in (\ref{L11}) with
\begin{align}
	(\varphi_2,\varphi_3)\sim(\varphi_2,\varphi_3)+\fft{2\pi}{p}(1-q,1+q)\,,
\end{align}
where $p$ and $q$ are coprime integers satisfying $0<q\leq p$ without loss of generality. Since we keep the same local metric (\ref{AdS5BH}) and simply generalize the periodic identification of coordinates, it is clear that this class of AdS$_5$ black holes with $\mathbb{R}\times L(p,q)$ conformal boundary solves the equations of motion of the 5d gauged supergravity STU model. See \cite{Cassani:2024kjn} for a brief discussion of a similar generalization in asymptotically flat 5d backgrounds.

\medskip

While the above generalization may appear straightforward, it carries an important holographic implication, paralleling the AdS$_7$ discussion in Section~\ref{sec:newBH:holo}. The supersymmetric limit of the Euclidean black hole solutions with $\mathbb{R}\times L(p,q)$ conformal boundaries discussed above can be uplifted to type IIB supergravity on $S^5$ and should provide the holographic dual description for the $S^1\times L(p,q)$ partition function, or lens space index \cite{Benini:2011nc,Nishioka:2014zpa}, of the 4d $\mathcal{N}=4$ SYM theory. This description should be valid in the large $N$ limit and ``on the second sheet'' \cite{Cassani:2021fyv,ArabiArdehali:2021nsx}. Moreover, the equal charge limit of the black hole solution is a valid background of the 5d $\mathcal{N}=2$ minimal gauged supergravity which describes the universal gravitational sector common to all 4d $\mathcal{N}=1$ holographic SCFTs. This minimal gauged supergravity lens space black hole can be uplifted in various ways to fully-fledged backgrounds of string or M-theory and should be a holographic dual of the corresponding 4d $\mathcal{N}=1$ lens space index ``on the second sheet''.\footnote{There could be topological restrictions imposed on the values of the integers $p$ and $q$ depending on the choice of internal manifold in the uplift to string or M-theory.} It will be interesting to understand this holographic duality in more detail since it provides a substantial generalization of earlier analyses in the $S^3 = L(1,1)$ case, see \cite{Hosseini:2017mds,Choi:2018hmj,Cabo-Bizet:2018ehj,Benini:2018ywd} and references thereof. Moreover, it is important to understand whether there is a smooth 5d supergravity background, akin to global AdS$_5$, which is the holographic dual to the lens space index ``on the first sheet''.

\section{Discussion}\label{sec:discussion}

In this paper we studied in detail the evaluation of the on-shell action of the BDHM solution of 7d gauged supergravity and showed that in the supersymmetric limit, and upon a particular choice of finite counterterms in the holographic renormalization procedure, the result agrees with the large $N$ limit of the superconformal index of the dual 6d $\mathcal{N}=(2,0)$ $A_N$ SCFT. We also discussed generalization of this 7d gauged supergravity background to new solutions with $\mathbb{R}\times L^{p,q,r}$ asymptotic boundary and $L^{p,q,r}$ horizons along with their holographic interpretation. We now briefly discuss some open questions and generalizations of our results that will be interesting to study in the future.

\begin{itemize}

\item In Section~\ref{sec:OSA} we showed how to evaluate the on-shell action of the BDHM background using holographic renormalization and a particular choice of finite counterterms that yields a result compatible with the first law of black hole thermodynamics and in agreement with holography. Nevertheless, it is clear that we have not provided a prescription for the possible finite counterterms that leads to satisfactory results for all known asymptotically locally AdS$_7$ backgrounds. It is very likely that, similar to the situation in 5d gauged supergravity discussed in \cite{BenettiGenolini:2016tsn}, one may in general need to invoke counterterms that break supersymmetry, gauge invariance or general covariance. This predicament could be understood from the perspective of the dual 6d SCFT where one will need to understand carefully the possible superconformal anomalies, as was done in \cite{Papadimitriou:2019gel,Katsianis:2019hhg} for 4d $\mathcal{N}=1$ SCFTs. It will be most interesting to investigate this subject further and arrive at a consistent holographic renormalization scheme that is applicable to general asymptotically AdS$_7$ backgrounds of holographic interest.

\item The black hole solution we constructed in Section~\ref{sec:newBH} has three equal angular momenta and an $L^{p,q,r}$ horizon. Since the $L^{p,q,r}$ SE manifolds are toric and thus have $\U(1)^3$ isometry, it could be expected that there is a generalization of this solution where the three angular momenta are not equal to each other. We have tried to find a suitable Ansatz for the 7d supergravity fields that leads to such a solution but were not able to solve the equations of motion. It will be interesting to either construct explicitly this hypothetical supergravity background or understand why it does not exist. This result will have interesting implications for the holographic dual 6d SCFT which will also be worth exploring. More generally, it will be very interesting to understand the constraints imposed by supergravity on the possible geometry and topology of 5d Riemannian manifolds that can appear as black hole horizons.

\item Our focus here was on the classical two-derivative on-shell action of the BDHM background, its supersymmetric limit and its relation to the dual supersymmetric SCFT partition function on $S^1\times S^5$ at leading order in the large $N$ limit. The first subleading correction to this supersymmetric free energy is known, see ~\cite{Nahmgoong:2019hko,Bobev:2015kza,Ohmori:2020wpk}, and it will be very interesting to reproduce this result using supergravity. To make progress on this challenging problem one would need to understand the leading higher-derivative corrections to 7d gauged supergravity and evaluate the on-shell action of this corrected supergravity action on the BDHM background. As suggested by recent holographic studies of higher-derivative corrections in 4d and 5d gauged supergravity, see \cite{Bobev:2020egg,Bobev:2021oku,Bobev:2021qxx,Bobev:2022bjm,Cassani:2022lrk}, it may be prudent to study this problem by first specializing to the minimal 7d gauged supergravity theory and therefore to the limit of three equal angular momenta and two equal charges in the BDHM background.

\item As stressed in Section~\ref{sec:newBH:holo}, the supersymmetric localization results for 6d SCFTs on $S^1\times L^{p,q,r}$ in \cite{Qiu:2013pta,Qiu:2014oqa} were analyzed in detail only on the ``first sheet'' in the terminology of~\cite{Cassani:2021fyv}. It will be interesting to revisit the analysis of this supersymmetric partition function focusing on the ``second sheet'' evaluation of the path integral and making connection with the analysis in~\cite{Ohmori:2020wpk} and~\cite{Cassani:2021fyv}. This will elucidate some field theoretic properties of the 6d SCFT and will allow for a more robust holographic comparison to our results for the on-shell action of the generalization of the supersymmetric BDHM solution to a background with $S^1\times L^{p,q,r}$ conformal boundary.

\item In Section~\ref{sec:newBH-5} we briefly discussed a simple generalization of the known AdS$_5$ black hole solutions of the 5d STU gauged supergravity model with $S^3$ horizons to similar backgrounds with $L(p,q)$ lens space horizons. These type of black holes are allowed by the general classification of possible horizon geometries and topologies in 5d gauged supergravity as discussed in \cite{Kunduri:2007qy}. Black holes with lens space horizon topology have been studied previously in 5d ungauged supergravity, see for example \cite{Kunduri:2014kja,Tomizawa:2016kjh} and \cite{Cassani:2025iix} for a recent discussion. Importantly, these are asymptotically flat black holes with a spatial $S^3$ at asymptotic infinity and a lens space horizon. In contrast, the solutions we present in Section~\ref{sec:newBH-5} have a lens space horizon and the same lens space topology at the AdS$_5$ boundary. It remains an open problem to understand whether more general black holes exist in AdS$_5$ for which the 3d spatial manifolds at the boundary and the horizon have lens space topology with different values of the integers $p$ and $q$.

\item Finally, we presented results for the on-shell actions of two particular, explicitly constructed, asymptotically $\text{AdS}_7$ solutions: a black hole with $S^1 \times S^5$ conformal boundary and a black hole with $S^1 \times L^{p,q,r}$ conformal boundary. It will be very interesting to extend the recent approach of equivariant localization, successfully applied to general supersymmetric saddles in asymptotically $\text{AdS}_4$ \cite{BenettiGenolini:2023kxp, Martelli:2023oqk} and asymptotically $\text{AdS}_5$ \cite{ Colombo:2025ihp, BenettiGenolini:2025icr} backgrounds, to the 7d $\text{U}\qty(1) \times \text{U}{(1)}$ gauged supergravity studied in this paper. In \cite{Colombo:2025ihp} it was shown that the on-shell action of a general supersymmetric saddle point of 5d minimal supergravity with $S^1 \times S^3$ conformal boundary is given by the Sasakian volume of a 5d compact manifold obtained from a 6d K\"ahler cone whose fan is the same as that of the original 5d supersymmetric saddle. It is tempting to speculate that, in a similar fashion, the on-shell action of a general 7d supersymmetric saddle with $S^1 \times S^5$ conformal boundary is related to the Sasakian volume of a 7d compact manifold obtained from 8d K\"ahler cone which has the same fan. It will also be interesting to learn how to generalize this statement to the supergravity backgrounds with $S^1 \times L^{p,q,r}$ case boundary. We leave these open questions for future work.

\end{itemize}

\section*{Acknowledgments}

We are grateful to Pieter Bomans, Davide Cassani, Fri\dh rik Freyr Gautason, Robie Hennigar, Dario Martelli, and Rishi Mouland for useful discussions. NB is supported in part by FWO projects G003523N, G094523N, and G0E2723N, as well as by the Odysseus grant G0F9516N from the FWO. JH is supported by the National Research Foundation of Korea(NRF) grant funded by the Korea government(MSIT) with grant number RS-2024-00449284, the Sogang University Research Grant of 202410008.01, the Basic Science Research Program of the National Research Foundation of Korea (NRF) funded by the Ministry of Education through the Center for Quantum Spacetime (CQUeST) with grant number RS-2020-NR049598. MD is supported by the Postdoctoral Fellowship of the Research Foundation- Flanders with the grant 1235324N. VD is partially supported by a grant Trapezio (2023) of the Fondazione Compagnia di San Paolo and would like to thank the ITF at KU Leuven for hospitality during various stages of this project.

\appendix

\section{Conventions}\label{app:conventions}
In a $D$-dimensional spacetime, the Hodge star operator acts on a $p$-form as
\begin{equation}
	\star(dx^{\mu_1}\wedge\cdots\wedge dx^{\mu_p})=\frac{1}{(D-p)!}\epsilon^{\mu_1\cdots \mu_p}{}_{\nu_{1}\cdots \nu_{D-p}}dx^{\nu_{1}}\wedge\cdots\wedge dx^{\nu_{D-p}}\,.\label{Hodge:dual}
\end{equation}
The totally anti-symmetric tensor is defined in the coordinate basis as
\begin{equation}
	\epsilon_{\mu_1\cdots\mu_D}=\begin{cases}
		\sqrt{|g|} & (\mu_1\cdots\mu_D~\text{is an even permutation of}~x^0\cdots x^{D-1}) \\
		-\sqrt{|g|} & (\mu_1\cdots\mu_D~\text{is an odd permutation of}~x^0\cdots x^{D-1}) \\
		0 & (\text{otherwise})
	\end{cases}\,,\label{epsilon}
\end{equation}
where $g$ denotes the determinant of the $D$-dimensional metric. Note that the definitions above apply both to mostly plus Lorentzian signature and to Euclidean signature.

\medskip

Consider the Wick rotation from a Lorentzian manifold to a Euclidean manifold in $D$-dimensions
\begin{align}
	X^0 \to -\ri X^0\qquad\&\qquad X_{0} \to \ri X_0\,, \label{Wick:general}
\end{align}
where we preserve the coordinate labeling by taking $X^0 = -\ri X^0$ rather than shifting to $X^0 = -\ri X^D$. Under the Wick rotation (\ref{Wick:general}), differential forms remain unaffected. However, the Hodge star operator (\ref{Hodge:dual}) transforms as
\begin{align}
	\star\omega_p \to -\ri\star \omega_p
\end{align}
as a consequence of the definition of the totally antisymmetric tensor (\ref{epsilon}), which applies in both Lorentzian and Euclidean signatures.

\section{Boundary Euclidean on-shell actions}\label{app:bdry}
In this Appendix, we provide technical details concerning the computation of the boundary Euclidean on-shell action, specifically the GHY term and the counterterms as discussed in Sections~\ref{sec:OSA:action} and~\ref{sec:OSA:on-shell}. 

\medskip

The first step is to derive explicit expressions for the integrands of the GHY term (\ref{GHY}) and the counterterms (\ref{ct-inf}) and (\ref{ct-fin}) evaluated on the $\tir=\tir_\infty$ boundary of the Euclidean BDHM background (\ref{BDHM:tilde}), subject to the asymptotic expansion (\ref{BDHM:asymp}). According to the analysis in Section~\ref{sec:OSA:on-shell}, this requires specifying the perturbation coefficients $\gamma^{(n)}_{\mu\nu}$ in the asymptotic expansion of the BDHM background (\ref{BDHM:asymp}), as well as the large $\tir$ behavior of the boundary curvature tensors. 

\medskip

Among the perturbation coefficients $\gamma^{(n)}_{\mu\nu}$, the first few leading terms relevant for the calculation of boundary Euclidean on-shell actions are given by
\begin{subequations}
	\begin{align}
		\gamma_{\tir\tir}^{(0)}&=-\fft{8m(s_1^2+s_2^2)}{5\mfg^2(\sum_{i=1}^3\Xi_i\tmu_i^2)^2}\,,\\
		\gamma_{\tir\tir}^{(1)}&=-\fft{2m(s_1^2+s_2^2)}{5\mfg^4(\sum_{i=1}^3\Xi_i\tmu_i^2)^2}-\fft{3m(s_1^2+s_2^2)\sum_{i=1}^3\Xi_i}{5\mfg^4(\sum_{i=1}^3\Xi_i\tmu_i^2)^3}+\fft{24m(s_1^2+s_2^2)\sum_{i=1}^3\Xi_i^2\tmu_i^2}{5\mfg^4(\sum_{i=1}^3\Xi_i\tmu_i^2)^4}\\
		&\quad+\fft{2m}{\mfg^4(\sum_{i=1}^3\Xi_i\tmu_i^2)^3}\bigg(1+\fft{s_1^2+s_2^2}{2}-\fft{(c_1-c_2)^2}{4}\bigg(2\Sigma_2+8\Pi_1+\Sigma_1\prod_{i=1}^3(\Sigma_1-2a_i\mfg)\bigg)\bigg)\,,\nn\\
		\gamma_{\tir a}^{(0)}&=-\fft{2m(s_1^2+s_2^2)}{\mfg^2(\sum_{i=1}^3\Xi_i\tmu_i^2)^3}(\Xi_a-\Xi_3)\tmu_a\qquad(\text{recall}~\tmu_3d\tmu_3=-\tmu_1d\tmu_1-\tmu_2d\tmu_2)\,,
	\end{align}\label{gamma}%
\end{subequations}
whereas the components exclusively along the 6d boundary coordinates $\gamma^{(n)}_{ij}=\alpha^{(n)}_{ij}$ are more intricate. For our purposes, we present only the terms necessary for evaluating the boundary actions:
\begin{subequations}
\begin{align}
	\alpha_{ij}dy^idy^j&=\fft{2m(s_1^2+s_2^2)}{5(\sum_{i=1}^3\Xi_i\tmu_i^2)^2}\bigg[(1+\mfg^2\tir_\infty^2)d\tau^2+\tir_\infty^2\sum_{i=1}^3\Big(d\tmu_i^2+\tmu_i^2d\phi_i^2\Big)\bigg]\fft{1}{\tir_\infty^4}+\mO(\tir_\infty^{-4})\,,\\
	\alpha&=\fft{12m(s_1^2+s_2^2)}{5(\sum_{i=1}^3\Xi_i\tmu_i^2)^2}\fft{1}{\tir_\infty^4}+\Bigg[-\fft{2m}{\mfg^2(\sum_{i=1}^3\Xi_i\tmu_i^2)^3}+\fft{2m(s_1^2+s_2^2)}{\mfg^2(\sum_{i=1}^3\Xi_i\tmu_i^2)^2}\\
	&\quad+\fft{12m(s_1^2+s_2^2)\sum_{i=1}^3\Xi_i}{5\mfg^2(\sum_{i=1}^3\Xi_i\tmu_i^2)^3}-\fft{46m(s_1^2+s_2^2)\sum_{i=1}^3\Xi_i^2\tmu_i^2}{5\mfg^2(\sum_{i=1}^3\Xi_i\tmu_i^2)^4}\nn\\
	&\quad-\fft{m\sum_{i=1}^3(2(s_1^2+s_2^2)-(c_1-c_2)^2(\Sigma_2-2a_i^2\mfg^2+\fft{2\Pi_1}{a_i^2\mfg^2}+2))(\Sigma_2-2a_i^2\mfg^2+\fft{2\Pi_1}{a_i^2\mfg^2})a_i^2\tmu_i^2}{2(\sum_{i=1}^3\Xi_i\tmu_i^2)^4}\nn\\
	&\quad+\fft{m(2(s_1^2+s_2^2)-(c_1-c_2)^2(\Sigma_2+2\Pi_1))(\Sigma_2+2\Pi_1-2)}{2\mfg^2(\sum_{i=1}^3\Xi_i\tmu_i^2)^4}\Bigg]\fft{1}{\tir_\infty^6}+\mO(\tir_\infty^{-8})\,.\nn
\end{align}\label{alpha}%
\end{subequations}
The boundary curvature tensors needed for evaluating the counterterms (\ref{ct-inf}) and (\ref{ct-fin}) admit the following large $\tir$ expansions
\begin{subequations}
	\begin{align}
		\sqrt{h}&=\sqrt{\bh}\bigg[1+\fft12\alpha+\mO(\tir_\infty^{-8})\bigg]\,,\\
		\mR&=\bmR+(-\alpha^{ij}\bmR_{ij}+\onabla_i\onabla_j\alpha^{ij}-\onabla_i\onabla^i\alpha)+\mO(\tir_\infty^{-8})\,,\\
		\mR_{ij}\mR^{ij}-\fft{3}{10}\mR^2&=\bmR_{ij}\bmR^{ij}-\fft{3}{10}\bmR^2+\mO(\tir_\infty^{-8})\,,\\
		\mR^3&=\bmR^3+\mO(\tir_\infty^{-10})\,,
	\end{align}\label{bdry-terms}%
\end{subequations}
which confirms that counterterms beyond the first two leading ones can be evaluated effectively using the global EAdS$_7$ background (\ref{EAdS7}) as argued in Section~\ref{sec:OSA:on-shell}. The leading non-trivial correction to the boundary Ricci scalar can be computed using the expressions~(\ref{alpha}) as
\begin{align}
	-\alpha^{ij}\bmR_{ij}+\onabla_i\onabla_j\alpha^{ij}-\onabla_i\onabla^i\alpha&=\Bigg[-\fft{8m(s_1^2+s_2^2)}{(\sum_{i=1}^3\Xi_i\tmu_i^2)^2}+\fft{16m(s_1^2+s_2^2)\sum_{i=1}^3\Xi_i}{(\sum_{i=1}^3\Xi_i\tmu_i^2)^3}\nn\\
	&\qquad-\fft{48m(s_1^2+s_2^2)\sum_{i=1}^3\Xi_i^2\tmu_i^2}{(\sum_{i=1}^3\Xi_i\tmu_i^2)^4}\Bigg]\fft{1}{\tir_\infty^6}+\mO(\tir_\infty^{-8})\,.\label{Ricci:correction}
\end{align}
The expressions presented above together with the global EAdS$_7$ information suffice to determine the integrands of the boundary Euclidean on-shell actions, the GHY term and the counterterms, for the Euclidean BDHM background (\ref{BDHM:tilde}).

\medskip

The next step involves evaluating the 6d boundary integrals with the integrands obtained above. For this, we first rewrite the unit 5-sphere metric from (\ref{EAdS7}) as
\begin{align}
	ds_{S^5}^2&=\fft{1-\tmu_2^2}{1-\tmu_1^2-\tmu_2^2}d\tmu_1^2+\fft{2\tmu_1\tmu_2}{1-\tmu_1^2-\tmu_2^2}d\tmu_1 d\tmu_2+\fft{1-\tmu_1^2}{1-\tmu_1^2-\tmu_2^2}d\tmu_2^2\nn\\
	&\quad+\tmu_1^2d\phi_1^2+\tmu_2^2d\phi_2^2+(1-\tmu_1^2-\tmu_2^2)d\phi_3^2\,.
\end{align}
With this, the 6d boundary integration measure becomes
\begin{align}
	\int d^6y\,\sqrt{\bh}(\cdots)=(2\pi)^3\beta\big(1+\mfg^2\tir^2\big)^{\fft12}\tir^5\int_{0\leq\tmu_1^2+\tmu_2^2\leq1}d\tmu_1 d\tmu_2\,\big(\tmu_1\tmu_2\big)(\cdots)\,,
\end{align}
where we have also used the coordinate ranges (\ref{BH:range}) and periodicity condition (\ref{tau:beta}). We then employ the following formulae ($i,j,k$ are all distinct)
\begin{subequations}
	\begin{align}
		\int_{0\leq\tmu_1^2+\tmu_2^2\leq1}d\tmu_1 d\tmu_2\,\big(\tmu_1\tmu_2\big)\fft{1}{(\sum_{i=1}^3\Xi_i\tmu_i^2)^2}&=-\fft14\sum_{i=1}^3\fft{\log\Xi_i}{(\Xi_i-\Xi_j)(\Xi_i-\Xi_k)}\,,\\
		\int_{0\leq\tmu_1^2+\tmu_2^2\leq1}d\tmu_1 d\tmu_2\,\big(\tmu_1\tmu_2\big)\fft{1}{(\sum_{i=1}^3\Xi_i\tmu_i^2)^3}&=\fft{1}{8\Xi_1\Xi_2\Xi_3}\,,\\
		\int_{0\leq\tmu_1^2+\tmu_2^2\leq1}d\tmu_1 d\tmu_2\,\big(\tmu_1\tmu_2\big)\fft{\tmu_i^2}{(\sum_{j=1}^3\Xi_j\tmu_j^2)^4}&=\fft{1}{24\Xi_1\Xi_2\Xi_3}\fft{1}{\Xi_i}\,,
	\end{align}\label{bdry:integral}%
\end{subequations}
which enable the explicit evaluation of the 6d boundary integrals. The final results for the GHY term (\ref{GHY}) and the counterterms (\ref{ct-inf}) and (\ref{ct-fin}) are presented in Section~\ref{sec:OSA:on-shell}.


\bibliography{AdS7BH}
\bibliographystyle{JHEP}

\end{document}